\documentclass[iop,apj]{emulateapj}
\usepackage{graphicx,url,afterpage}

\shorttitle{Mass--Metallicity Relation for Dwarf Galaxies}
\shortauthors{Kirby et al.}
\slugcomment{Accepted to the Astrophysical Journal on Tuesday, October 1, 2013}

\newcommand{\lfoffsetdsph}{-1.69}
\newcommand{\lfslopedsph}{0.29}
\newcommand{\lfoffseterrdsph}{0.06}
\newcommand{\lfslopeerrdsph}{0.04}
\newcommand{\lfoffsetboth}{-1.68}
\newcommand{\lfslopeboth}{0.29}
\newcommand{\lfoffseterrboth}{0.03}
\newcommand{\lfslopeerrboth}{0.02}
\newcommand{\lfoffsetdirr}{-1.58}
\newcommand{\lfslopedirr}{0.21}
\newcommand{\lfoffseterrdirr}{0.04}
\newcommand{\lfslopeerrdirr}{0.02}

\newcommand{\mfoffsetboth}{-1.69}
\newcommand{\mfslopeboth}{0.30}
\newcommand{\mfoffseterrboth}{0.04}
\newcommand{\mfslopeerrboth}{0.02}

\newcommand{\lfrmsdsph}{0.17}
\newcommand{\lfrmsdirr}{0.09}

\newcommand{\lfrmsdirrdsph}{0.12}
\newcommand{\lfrms}{0.16}
\newcommand{\mfrms}{0.17}
\newcommand{\forfeh}{-1.04}
\newcommand{\nsettfeh}{-1.05}
\newcommand{\nofsfeh}{-0.83}
\newcommand{\noeffeh}{-1.12}
\newcommand{\ntzffeh}{-0.92}

\newcommand{\leoiim}{ 3.3^{+0.5}_{-0.7}}
\newcommand{\umim}{11.0^{+4.5}_{-5.6}}
\newcommand{\nsettm}{ 1.7^{+0.2}_{-0.3}}
\newcommand{\ksdsphdirr}{0.02}

\begin{document}
    % See p.105 of "TeX Unbound" for suggested values.
    % See pp. 199-200 of Lamport's "LaTeX" book for details.
    %   General parameters, for ALL pages:
    \renewcommand{\topfraction}{0.9}	% max fraction of floats at top
    \renewcommand{\bottomfraction}{0.9}	% max fraction of floats at bottom
    %   Parameters for TEXT pages (not float pages):
    \setcounter{topnumber}{6}
    \setcounter{bottomnumber}{6}
    \setcounter{totalnumber}{6}     % 2 may work better
    \setcounter{dbltopnumber}{6}    % for 2-column pages
    \renewcommand{\dbltopfraction}{0.9}	% fit big float above 2-col. text
    \renewcommand{\textfraction}{0.1}	% allow minimal text w. figs
    %   Parameters for FLOAT pages (not text pages):
    \renewcommand{\floatpagefraction}{0.8}	% require fuller float pages
	% N.B.: floatpagefraction MUST be less than topfraction !!
    \renewcommand{\dblfloatpagefraction}{0.8}	% require fuller float pages
\newcommand{\mathfeh}{{\rm [Fe/H]}}

\title{The Universal Stellar Mass--Stellar Metallicity Relation for Dwarf Galaxies}

\author{Evan~N.~Kirby\altaffilmark{1,2},
  Judith~G.~Cohen\altaffilmark{3}, 
  Puragra~Guhathakurta\altaffilmark{4},
  Lucy~Cheng\altaffilmark{5,6},
  James~S.~Bullock\altaffilmark{1},
  \and Anna~Gallazzi\altaffilmark{7,8}}

\altaffiltext{*}{The data presented herein were obtained at the
  W.~M.~Keck Observatory, which is operated as a scientific
  partnership among the California Institute of Technology, the
  University of California and the National Aeronautics and Space
  Administration. The Observatory was made possible by the generous
  financial support of the W.~M.~Keck Foundation.}
\altaffiltext{1}{University of California, Department of Physics and
  Astronomy, 4129 Frederick Reines Hall, Irvine, CA 92697, USA,
  ekirby@uci.edu}
\altaffiltext{2}{Center for Galaxy Evolution Fellow.}
\altaffiltext{3}{California Institute of Technology, Department of
  Astronomy \& Astrophysics, 1200 E.\ California Blvd., MC 249-17,
  Pasadena, CA 91125, USA}
\altaffiltext{4}{UCO/Lick Observatory and Department of Astronomy and
  Astrophysics, University of California, Santa Cruz, CA 95064, USA}
\altaffiltext{5}{The Harker School, 500 Saratoga Ave., San Jose, CA
  95117, USA}
\altaffiltext{6}{Harvard University, Department of Astronomy, 60
  Garden St., Cambridge, MA 02138, USA}
\altaffiltext{7}{INAF -- Osservatorio Astrofisico di Arcetri, Largo
  E. Fermi 5, I-50125 Firenze, Italy}
\altaffiltext{8}{Dark Cosmology Centre, Niels Bohr Institute,
  University of Copenhagen, Juliane Mariesvej 30, 2100 Copenhagen,
  Denmark}


%%%%%%%%%%%%%%%%%%%%%%%%%%%%%%%%%
%%%%%%%%%    ABSTRACT    %%%%%%%%
%%%%%%%%%%%%%%%%%%%%%%%%%%%%%%%%%

\begin{abstract}

We present spectroscopic metallicities of individual stars in seven
gas-rich dwarf irregular galaxies (dIrrs), and we show that dIrrs obey
the same mass--metallicity relation as the dwarf spheroidal (dSph)
satellites of both the Milky Way and M31: $Z_* \propto
M_*^{\mfslopeboth \pm \mfslopeerrboth}$.  The uniformity of the
relation is in contradiction to previous estimates of metallicity
based on photometry.  This relationship is roughly continuous with the
stellar mass--stellar metallicity relation for galaxies as massive as
$M_* = 10^{12}~M_{\sun}$.  Although the average metallicities of dwarf
galaxies depend only on stellar mass, the shapes of their metallicity
distributions depend on galaxy type.  The metallicity distributions of
dIrrs resemble simple, leaky box chemical evolution models, whereas
dSphs require an additional parameter, such as gas accretion, to
explain the shapes of their metallicity distributions.  Furthermore,
the metallicity distributions of the more luminous dSphs have sharp,
metal-rich cut-offs that are consistent with the sudden truncation of
star formation due to ram pressure stripping.

\end{abstract}

%%%%%%%%%%%%%%%%%%%%%%%%%%%%%%%%%
%%%%%%%%%   SECTION 1   %%%%%%%%%
%%%%%%%%%%%%%%%%%%%%%%%%%%%%%%%%%

\section{Introduction}
\label{sec:intro}

The average metal content of a galaxy correlates with its mass.  More
massive galaxies are more metal-rich than less massive galaxies.  The
relation can be explained by the retention of metals in the galaxies'
gravitational potential wells \citep[e.g.,][]{dek86}.  High-mass
galaxies have deep potential wells that can resist some of the
expulsion of gas and metals by supernova winds, stellar winds, and
galaxy-scale feedback.  Low-mass galaxies lack the gravity to resist
these feedback mechanisms.  The correlation between metallicity and
mass can also be explained by a correlation between star formation
efficiency and stellar mass \citep[e.g.,][]{mat94,cal09,mag12,pip13}.
If massive galaxies evolve quickly, then they can achieve high stellar
masses and low gas mass fractions.  Consequently, their metallicities
will be high.  On the other hand, slowly evolving, low-mass galaxies
can have high gas fractions, which dilute the metallicity of both the
gas and the stars that form from the gas.  Yet another explanation for
the mass--metallicity relation is a stellar initial mass function
(IMF) that changes with the rate of star formation \citep{kop07}.
Because that rate depends on galaxy mass and because the metal yield
depends on the masses of stars, a galaxy's metallicity then depends on
its stellar mass.

Galactic metallicity is typically measured in the gas phase.  Emission
line diagnostics of metallicity \citep[e.g.,][]{kew02} sample the
metallicity of \ion{H}{2} regions.  Hence, these measures probe
presently star-forming gas.  Using strong line diagnostics,
\citet{mcc68} and \citet{leq79} established the first
luminosity--metallicity relations (LZRs) for star-forming galaxies.
\citeauthor{leq79}\ found that the effective metal yields of all of
the galaxies they considered were below the true yield expected from a
``closed box'' or ``simple'' model of galactic chemical evolution
\citep{sch63,tal71,sea72}.  More recently, \citet{tre04} showed that
the average metallicities of star-forming galaxies in the Sloan
Digital Sky Survey \citep[SDSS,][]{aba04} correlate strongly with
their stellar masses or rotation speeds.  The correlation with stellar
mass has an intrinsic scatter of only 0.1~dex in $\log ({\rm O/H})$.
As with previous studies, \citeauthor{tre04}\ interpreted the relation
as a progression of a larger effective yield for more massive
galaxies.  Expressed another way, more massive galaxies lose a smaller
fraction of the metals that their stars produce.

The mass--metallicity relation (MZR)---where metallicity was measured
in the gas phase---extends down to the mass range of dwarf galaxies as
small as a few million solar masses.  \citet{mou83} and \citet{ski89}
showed that dwarf elliptical (dE) and dwarf irregular galaxies (dIrrs)
in and around the Local Group (LG) obey a LZR\@.  \citet{gar02}
extended the relation to more distant spiral galaxies.  Irregular and
spiral galaxies obey the same, unbroken relation over 4.5 orders of
magnitude in luminosity.  %The
%relationship is linear below a stellar mass of $M_* =
%10^{10}~M_{\sun}$, and the slope decreases above that mass.

One of the sources of error in determining the MZR is the stellar
mass-to-light ratio ($M_*/L$).  This ratio is required to convert the
LZR into the MZR\@.  It depends on a galaxy's star formation history
(SFH) with potential contributions from the stellar IMF\@.
\citet{lee06a} measured stellar masses for dwarf galaxies based on
4.5~$\mu$m luminosity, which is less sensitive to age, SFH, and dust
than visible luminosity.  While \citet{ski89} already established that
the LZR applies to dwarf galaxies, \citet{lee06a} showed that the MZR
from more massive spiral galaxies \citep{tre04} also applies to dIrrs.

The MZR also exists at high redshift.  \citet{erb06} found that the
relation persists, but evolves in the sense that high-redshift
galaxies are more metal-poor at a given stellar mass than galaxies in
the local universe.  \citet{zah13} and \citet{hen13} showed that the
MZR evolves smoothly from $z=2.2$ to the present.  However,
\citet{man10}, \citet{hun12}, and \citet{lar13} found that the
independent variable controlling the offset of the MZR was star
formation rate (SFR), not redshift.  Because SFR increases with
redshift \citep{mad96}, it appeared as though the MZR was evolving,
but it is in fact constant after the correction for SFR\@.  Correcting
for SFR leads to the unevolving ``fundamental metallicity relation.''
Star formation depresses the gas-phase metallicity of a galaxy because
star formation requires hydrogen gas.  Metallicity is expressed as the
ratio of metals to hydrogen.  Therefore, vigorously star forming
galaxies contain a lot of hydrogen, which dilutes the metals.  In
fact, \citet{bot13} established that the offset in the MZR correlates
better with \ion{H}{1} gas mass than SFR\@.

The majority of stellar mass in the local universe resides in galaxies
that are not forming stars and have very little gas
\citep{bel03,bal04,gal08}.  Therefore, it is not possible to place
them on the same fundamental metallicity relation corrected for SFR or
gas mass.  Instead, it is necessary to construct a separate MZR that
measures stellar metallicity rather than gas-phase metallicity.

Emission line diagnostics of metallicity apply only to star-forming
galaxies.  They sample the present metallicity of stars forming at the
time of observation.  A complementary technique for measuring the
composition of a galaxy is to measure stellar metallicities from
stellar colors or spectral absorption lines.  \citet{bau59} first
noticed a correlation between galaxies' $B-V$ colors and their
absolute magnitudes.  He suggested that the cause of the correlation
was a variable ratio of Population~I (young, metal-rich) to
Population~II (old, metal-poor) stars.  In essence, he first suggested
the idea of a LZR\@.  Later, \citet{san78} showed that the
color--magnitude relation applies to both Virgo cluster galaxies and
field galaxies.

The stellar metallicity of a galaxy is a record of the past star
formation.  Each star preserves the metallicity of the galactic gas at
the time and site of formation.  A stellar metallicity distribution
function is therefore a chronicle of the chemical evolution of the
galaxy.  In the typical mode of a monotonic increase of gas
metallicity with time, the gas-phase metallicity is greater than the
average stellar metallicity.  Furthermore, the gas-phase metallicity
in principle fluctuates more than the stellar metallicity.  The gas
metallicity changes as rapidly as gas flows into and out of the
galaxy, whereas the stellar metallicity responds to gas flows on a
longer timescale, which depends on the SFR\@.  On the other hand,
\citet{ber11,ber12} showed that the dispersion in the MZR is just
0.15~dex when the ``direct'' method is used to measure gas-phase
abundances from auroral lines.  Furthermore, rare outliers from the
MZR when using the strong nebular lines are not outliers when using
the more trusted, direct method.  The low dispersion leaves little
room for large fluctuations in the gas-phase metallicity.  However,
the direct method is only practical for a limited range of galaxies
because the [\ion{O}{3}]~$\lambda$4363 emission line is intrinsically
faint, especially at high metallicities ($12 + \log ({\rm O/H}) \ga
8.5$).

Quiescent and gas-free galaxies have no emission lines.  The only
light from these galaxies comes from stars.  Hence, spectroscopic
metallicities must be measured from absorption lines rather than
emission lines.  In practice, models of the integrated light spectrum
from an entire stellar population \citep{tin76} are compared to
observed galaxy spectra.  One implementation of this technique is to
measure spectrophotometric indices, such as Lick indices
\citep{fab73,wor94}.  \citet{bru03} updated the use of line indices to
rely on spectral models rather than fixed-resolution templates.  This
method can even be used with ultraviolet line indices for
high-redshift galaxies \citep{rix04,hal08,som12}.  Spectral models can
even be compared directly to observed spectra without compressing the
abundance information into a few line indices \citep[e.g.,][]{con09}.
\citet{gal05} applied the \citeauthor{bru03}\ models to SDSS spectra
of over 40,000 galaxies.  They showed that the average stellar
metallicities of galaxies with stellar masses in the range $10^9 <
M_*/M_{\sun} < 10^{12}$ correlate tightly with stellar mass.  The
correlation has the same shape as the correlation between gas-phase
metallicity and stellar mass for SDSS galaxies found by \citet{tre04}.
\citet{gal06} showed that early-type galaxies' stellar metallicities
correlate better with their dynamical masses than their stellar
masses.  The correlation with dynamical mass lends credence to the
theory that galaxies with shallower potential wells are more
susceptible to metal loss.

On the other hand, dark matter halo masses for dwarf galaxies ($M_* <
10^8~M_{\sun}$) may not correlate at all with their stellar masses or
metallicities \citep{str08}.  However, the full dark matter potential
is difficult to measure in any galaxy.  While it is straightforward to
constrain the mass within the half-light radius \citep{wol10}, it is
nearly impossible to measure the total gravitational potentials of
most dispersion-supported galaxies in the absence of an extended
tracer \citep{tol11}.  The baryons are so deeply embedded in the dark
matter halo that any connection to halo virial mass requires a
theoretical extrapolation.

Gas-phase abundances are usually expressed in terms of oxygen
abundance.  Oxygen absorption lines in stars are few and weak.  The
spectral features most readily available in stars are from iron and
magnesium.  Therefore, comparing gas-phase to stellar metallicities
requires the assumption of abundance ratios, such as [O/Fe]\@.  This
ratio depends on the SFH\@.  Another option to compare gas-phase
abundances in star-forming galaxies to quiescent galaxies is to
measure oxygen abundances of planetary nebulae (PNe), which are the
long-lived remnants of dead stars.  \citet{ric95} measured the oxygen
abundances of PNe in both dIrrs and dwarf elliptical/spheroidal
galaxies (dEs/dSphs).  They found that the dE/dSph LZR was offset from
dIrrs in the sense that PNe in dEs and dSphs are more oxygen-rich than
in dIrrs of similar luminosity.  There is some question about whether
oxygen abundances in PNe trace the oxygen abundances of stars.
\citet{ric98} laid out the reasons for possible discrepancies, but did
not find them to apply to the PNe in dwarf galaxies.  \citet{gon07}
revisited the offset, and they showed that it applies whether the
oxygen abundances in dIrrs are measured from PNe or \ion{H}{2}
regions.  One possibility is that the PNe themselves produce oxygen in
the third dredge-up, while they are expelling their envelopes
\citep[e.g.,][]{mag05}.  Another possibility is that PNe and
\ion{H}{2} regions could preferentially trace a younger, more
metal-rich population than the average.

\begin{figure}[t!]
\centering
\includegraphics[width=\columnwidth]{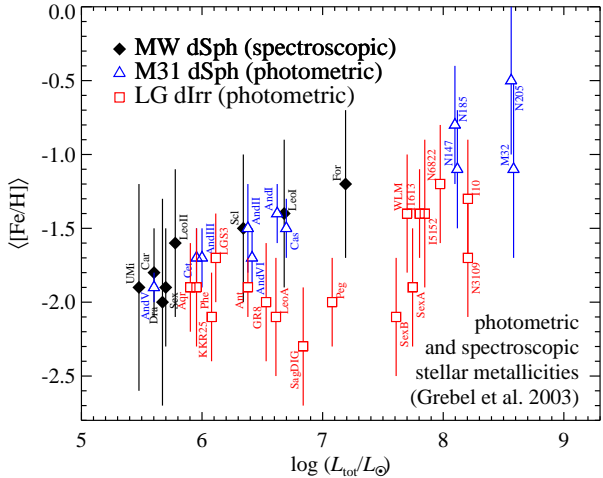}
\caption{The luminosity--stellar metallicity relation for Local Group
  dwarf galaxies where the metallicity is determined from spectroscopy
  (filled symbols) and photometry (hollow symbols).  The data is taken
  from the compilation of \citet{gre03}.  There is an apparent offset
  between dIrrs and dSphs.  We argue that the offset is not a
  reflection of the dIrrs' true metallicities and is instead caused by
  the age--metallicity degeneracy for photometric
  metallicities.\label{fig:gre03}}
\end{figure}

Nearby galaxies can be resolved into individual \ion{H}{2} regions and
individual stars.  Whereas integrated light spectra do not allow a
measurement of the spread of metallicity within a galaxy, spectroscopy
of individual stars resolves the shapes of metallicity distributions.
%For gas-phase abundances, the ability to resolve \ion{H}{2} regions
%spatially revealed that the oxygen abundance decreases with
%increasing radius from the center of the galaxy \citep{sea71}.
Individual stellar metallicities can be measured from color--magnitude
diagrams (CMDs).  \citet{gre03} compiled average metallicities of LG
dwarf galaxies.  The metallicities were measured spectroscopically for
the nearer dSph galaxies and from the optical colors of the red giant
branch (RGB) for the more distant dIrr galaxies.  In agreement with
the oxygen measurements from PNe, \citet{gre03} found an offset in the
LZR such that dIrrs are more iron-poor at fixed luminosity than dSphs
(see Figure~\ref{fig:gre03}).

However, the colors of red giants are subject to the age--metallicity
degeneracy \citep[e.g.,][]{sal05,lia11}.  As a stellar population
ages, the RGB becomes redder.  However, more metal-rich red giants are
also redder.  Thus, the age of the population needs to be determined
before the RGB color can be used to measure metallicity.  Determining
the age is especially important for a comparison between dSphs and
dIrrs because the stellar populations of dSphs are systematically
older than dIrrs.  \citet{lee08} reported in conference proceedings
that they re-analyzed the metallicities of dIrrs from RGB colors, and
they found iron metallicities on average 0.5~dex higher than
\citet{gre03} for the same galaxies.  The difference in the analyses
arose from a different treatment in the ages of the stellar
populations as well as accounting for a spread in age and metallicity.
The offset between dIrrs and dSphs in the LZR vanished in the more
recent study.

When it is possible to observe stars spectroscopically, spectroscopic
metallicities are preferred to photometric metallicities because the
age--metallicity degeneracy applies only in subtle ways to stellar
spectra.  \citet{arm91} first measured the spectroscopic metallicities
for extragalactic stars.  They based their measurements of the Carina
dSph on the strength of the near-infrared \ion{Ca}{2} triplet (CaT),
which they calibrated to Galactic globular clusters of known
metallicity.  Since then, the CaT technique has been used to quantify
the metallicity distributions of most of the classical dSph Milky Way
(MW) satellites \citep[e.g.,][]{hel06} and two dIrrs: NGC~6822
\citep{tol01} and WLM at a distance of 930~kpc \citep{lea09}.  Using a
technique not based on empirical calibrations, \citet{kir08b,kir11a}
quantified the metallicity distributions of 15 MW dSphs using
synthetic spectral fitting to iron lines in individual red giants.
They found a single, power-law relation between metallicity and galaxy
luminosity over the range $10^{3.5} < L/L_{\sun} < 10^{7.3}$.  This
relation for dSphs is consistent with the MZR determined from
photometric metallicities of stars in more massive dEs
\citep{gre03,woo08}.

In the present study, we extend our spectroscopic analysis of iron
lines in red giants to gas-rich dIrrs.  We aim to resolve the
ambiguity of the photometric metallicity measurements in dIrrs.  We
construct a unified stellar mass--stellar metallicity relation for LG
dwarf galaxies of various morphologies, ages, and gas fractions.  This
relation spans six orders of magnitude in luminosity or stellar mass.
We further connect this relation to the stellar mass--stellar
metallicity relation of more massive galaxies \citep{gal05}.  We
discuss the origin of the universal MZR in the context of metal loss.
Finally, we present the shapes of the metallicity distributions and
discuss the role of gas inflow, outflow, and ram pressure stripping.

%\afterpage{\clearpage}

%%%%%%%%%%%%%%%%%%%%%%%%%%%%%%%%%
%%%%%%%%%   SECTION 2   %%%%%%%%%
%%%%%%%%%%%%%%%%%%%%%%%%%%%%%%%%%

\section{Spectroscopic Observations}
\label{sec:obs}

\subsection{The Galaxy Sample}
\label{sec:galaxies}

\begin{deluxetable*}{lccccccc}
\tablewidth{0pt}
\tablecolumns{8}
\tablecaption{Dwarf Galaxy Sample\label{tab:sample}}
\tablehead{\colhead{Galaxy} & \colhead{RA (J2000)} & \colhead{Dec (J2000)} & \colhead{$\log M_*$} & \colhead{$D_{\rm MW}$} & \colhead{$D_{\rm M31}$} & \colhead{$v_{\rm rot}/\sigma_v$}\tablenotemark{a} & \colhead{$M_{\rm HI}/M_*$} \\
\colhead{ } & \colhead{ } & \colhead{ } & \colhead{($M_{\sun}$)} & \colhead{(kpc)} & \colhead{(kpc)} & \colhead{ } & \colhead{ }}
\startdata
\cutinhead{Milky Way dSphs}
Fornax                       & 02 39 59 & $-$34 26 57 & 7.39 &    \phn149 &  \nodata   & $<1$    & 0\tablenotemark{b} \\
Leo I                        & 10 08 28 & $+$12 18 23 & 6.69 &    \phn258 &  \nodata   & $<1$    & 0 \\
Sculptor                     & 01 00 09 & $-$33 42 33 & 6.59 & \phn\phn86 &  \nodata   & $<1$    & 0\tablenotemark{b} \\
Leo II                       & 11 13 29 & $+$22 09 06 & 6.07 &    \phn236 &  \nodata   & \nodata & 0 \\
Sextans                      & 10 13 03 & $-$01 36 53 & 5.84 & \phn\phn89 &  \nodata   & $<1$    & 0 \\
Ursa Minor                   & 15 09 08 & $+$67 13 21 & 5.73 & \phn\phn78 &  \nodata   & $<1$    & 0 \\
Draco                        & 17 20 12 & $+$57 54 55 & 5.51 & \phn\phn76 &  \nodata   & $<1$    & 0 \\
Canes Venatici I             & 13 28 04 & $+$33 33 21 & 5.48 &    \phn218 &  \nodata   & \nodata & 0 \\
Hercules                     & 16 31 02 & $+$12 47 30 & 4.57 &    \phn126 &  \nodata   & \nodata & 0 \\
Ursa Major I                 & 10 34 53 & $+$51 55 12 & 4.25 &    \phn102 &  \nodata   & \nodata & 0 \\
Leo IV                       & 11 32 57 & $-$00 32 00 & 3.93 &    \phn155 &  \nodata   & \nodata & 0 \\
Canes Venatici II            & 12 57 10 & $+$34 19 15 & 3.90 &    \phn161 &  \nodata   & \nodata & 0 \\
Ursa Major II                & 08 51 30 & $+$63 07 48 & 3.73 & \phn\phn38 &  \nodata   & $<1$    & 0 \\
Coma Berenices               & 12 26 59 & $+$23 54 15 & 3.68 & \phn\phn45 &  \nodata   & $<1$    & 0 \\
Segue 2                      & 02 19 16 & $+$20 10 31 & 3.01 & \phn\phn41 &  \nodata   & \nodata & \nodata \\
\cutinhead{Local Group dIrrs}
NGC 6822                     & 19 44 56 & $-$14 47 21 & 7.92 &    \phn452 &    \phn897 & 8.1     & 1.6 \\
IC 1613                      & 01 04 48 & $+$02 07 04 & 8.01 &    \phn758 &    \phn520 & 4.2     & 0.6 \\
VV 124\tablenotemark{c}      & 09 16 02 & $+$52 50 24 & 7.00 &       1367 &       1395 & $<1$    & 0.1 \\
Pegasus dIrr\tablenotemark{c} & 23 28 36 & $+$14 44 35 & 6.82 &    \phn921 &    \phn474 & $>2.3$  & 0.9 \\
Leo A                        & 09 59 27 & $+$30 44 47 & 6.47 &    \phn803 &       1200 & $<1$    & 3.7 \\
Aquarius                     & 20 46 52 & $-$12 50 53 & 6.15 &       1066 &       1173 & $<1$    & 2.9 \\
Leo T\tablenotemark{c}       & 09 34 53 & $+$17 03 05 & 5.21 &    \phn422 &    \phn991 & $<1$    & 1.7 \\
\cutinhead{M31 dSphs}
NGC 205\tablenotemark{d}     & 00 40 22 & $+$41 41 07 & 8.67 &  \nodata   & \phn\phn42 & 0.3     & 0.0009 \\
NGC 185\tablenotemark{d}     & 00 38 58 & $+$48 20 15 & 7.83 &  \nodata   &    \phn187 & 0.6     & 0.0016 \\
NGC 147\tablenotemark{d}     & 00 33 12 & $+$48 30 32 & 8.00 &  \nodata   &    \phn142 & 1.1     & 0 \\
Andromeda VII                & 23 26 32 & $+$50 40 33 & 6.93 &  \nodata   &    \phn218 & \nodata & 0 \\
Andromeda II                 & 01 16 30 & $+$33 25 09 & 6.88 &  \nodata   &    \phn184 & \nodata & 0 \\
Andromeda I                  & 00 45 40 & $+$38 02 28 & 6.80 &  \nodata   & \phn\phn58 & \nodata & 0 \\
Andromeda III                & 00 35 34 & $+$36 29 52 & 6.18 &  \nodata   & \phn\phn75 & \nodata & 0 \\
Andromeda XVIII\tablenotemark{e} & 00 02 15 & $+$45 05 20 & 5.78 &       1358 &    \phn591 & \nodata & \nodata \\
Andromeda XV                 & 01 14 19 & $+$38 07 03 & 5.76 &  \nodata   &    \phn174 & \nodata & 0 \\
Andromeda V                  & 01 10 17 & $+$47 37 41 & 5.63 &  \nodata   &    \phn110 & \nodata & 0 \\
Andromeda XIV                & 00 51 35 & $+$29 41 49 & 5.45 &  \nodata   &    \phn162 & \nodata & 0 \\
Andromeda IX                 & 00 52 53 & $+$43 11 45 & 5.26 &  \nodata   & \phn\phn40 & \nodata & 0 \\
Andromeda X                  & 01 06 34 & $+$44 48 16 & 5.02 &  \nodata   &    \phn110 & \nodata & 0 \\
\enddata
\tablerefs{The June 2013 version of the compilation of \citet{mcc12} and references therein.}
\tablenotetext{a}{The ratio of the rotation velocity to the velocity dispersion.  For the MW and M31 dSphs, the ratio is measured from stellar velocities.  For the LG dIrrs, the ratio is measured from \ion{H}{1} gas velocities.  Following the advice of \citet{mcc12}, we assume that the upper limit on the rotation velocity is the velocity dispersion in the cases where no rotation was detected.  This gives an upper limit of $v_{\rm rot}/\sigma_v < 1$.  The ratio is not given in cases where measurements are missing or where neither rotation nor dispersion has been detected.}
\tablenotetext{b}{Some \ion{H}{1} has been detected along the line of sight to these galaxies \citep{car98,bou06}, but it is probably not associated with these galaxies \citep{grc09}.}
\tablenotetext{c}{Transition dwarf galaxies, alternately notated as dTs or dIrr/dSphs.}
\tablenotetext{d}{These galaxies are traditionally classified as dEs, but they have the same properties as high-luminosity dSphs (see Section~\ref{sec:galaxies}).}
\tablenotetext{e}{\citet{mcc08} and \citet{mcc12} classified Andromeda~XVIII as an isolated dSph, but we list it in the M31 system anyway.}
\end{deluxetable*}

\begin{figure*}[p]
\centering
\includegraphics[width=0.46\textwidth]{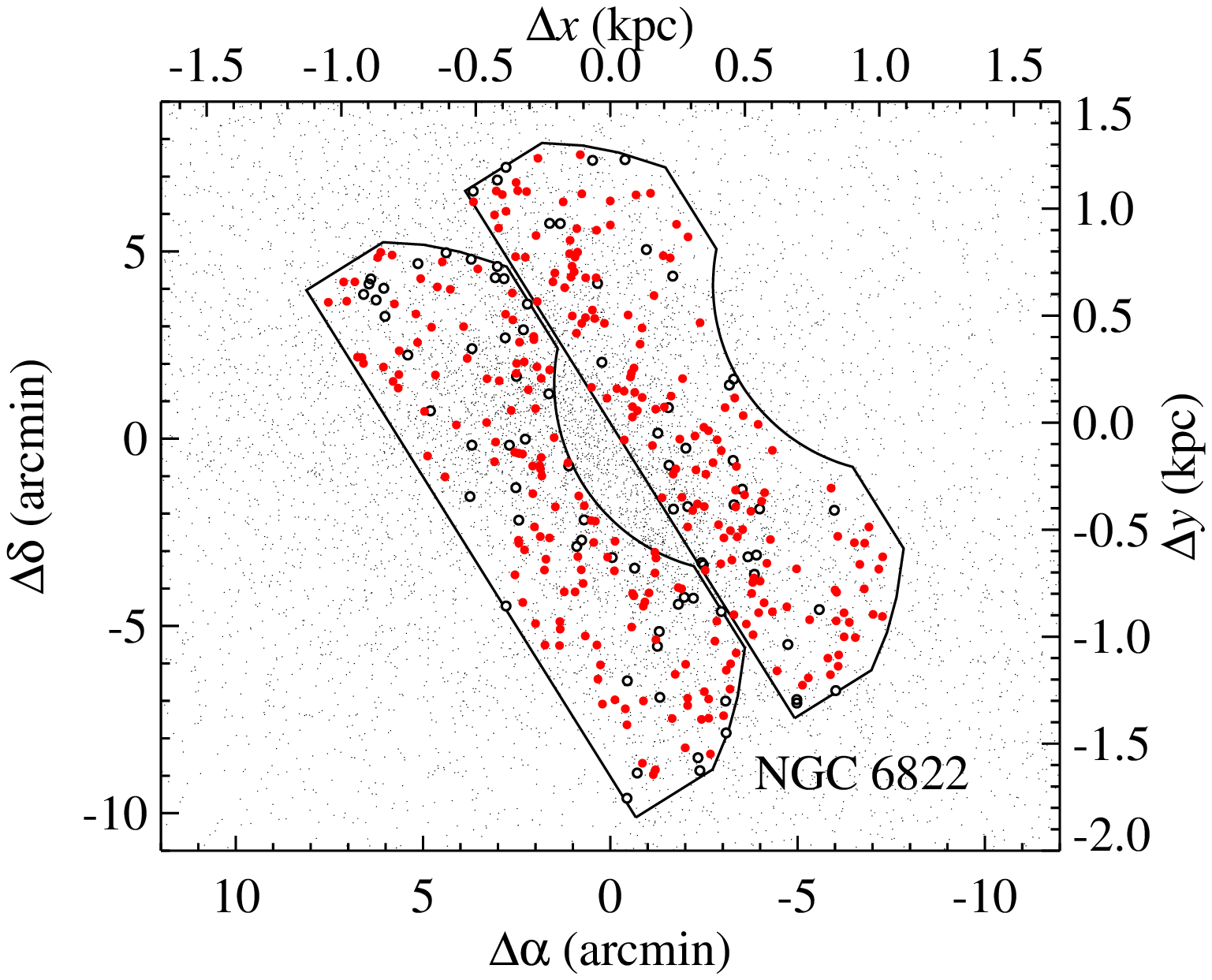}
\hfil
\includegraphics[width=0.46\textwidth]{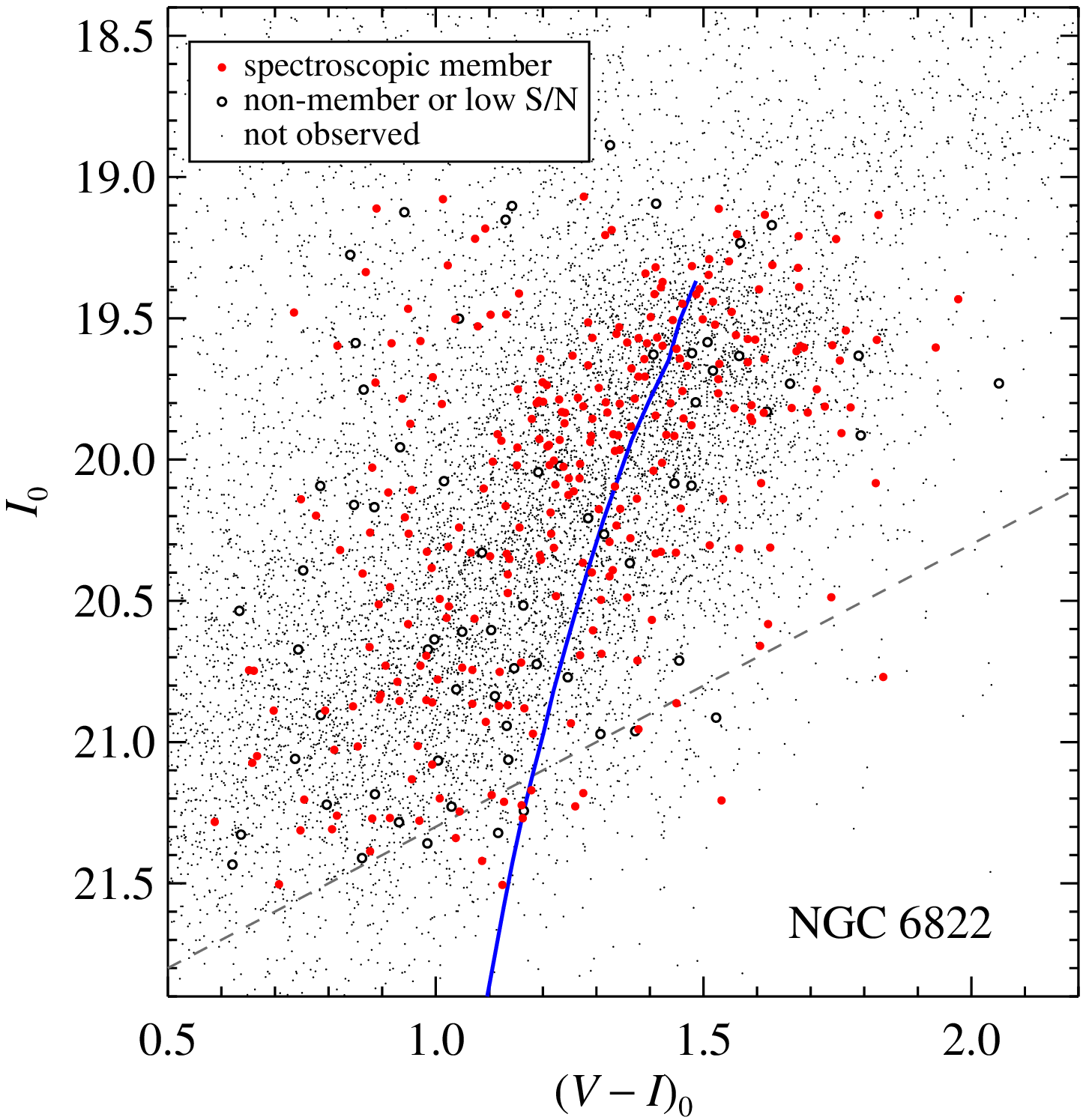}
\caption{Stars from the NGC~6822 photometric catalog \protect
  \citep{mas07} represented in celestial coordinates (left) and in a
  color--magnitude diagram (right).  The irregular shapes in the left
  panel show the outlines of the DEIMOS field of view for both
  slitmasks.  Solid red and hollow black points show targets for which
  we obtained DEIMOS spectra.  Solid red points are spectroscopically
  confirmed members.  The blue curve in the right panel is a Padova
  theoretical isochrone \protect \citep{gir02} with an age of 6~Gyr
  and a metallicity of $\mathfeh = -1.2$.  The age of the isochrone is
  not meant to indicate the true age of the galaxy.  It was chosen
  merely to illustrate the approximate shape of the RGB\@.  Stars
  below the dashed gray line have uncertainties in $V_0$ larger than
  0.1~dex.\label{fig:n6822}}
\end{figure*}

\begin{figure*}[p]
\centering
\includegraphics[width=0.46\textwidth]{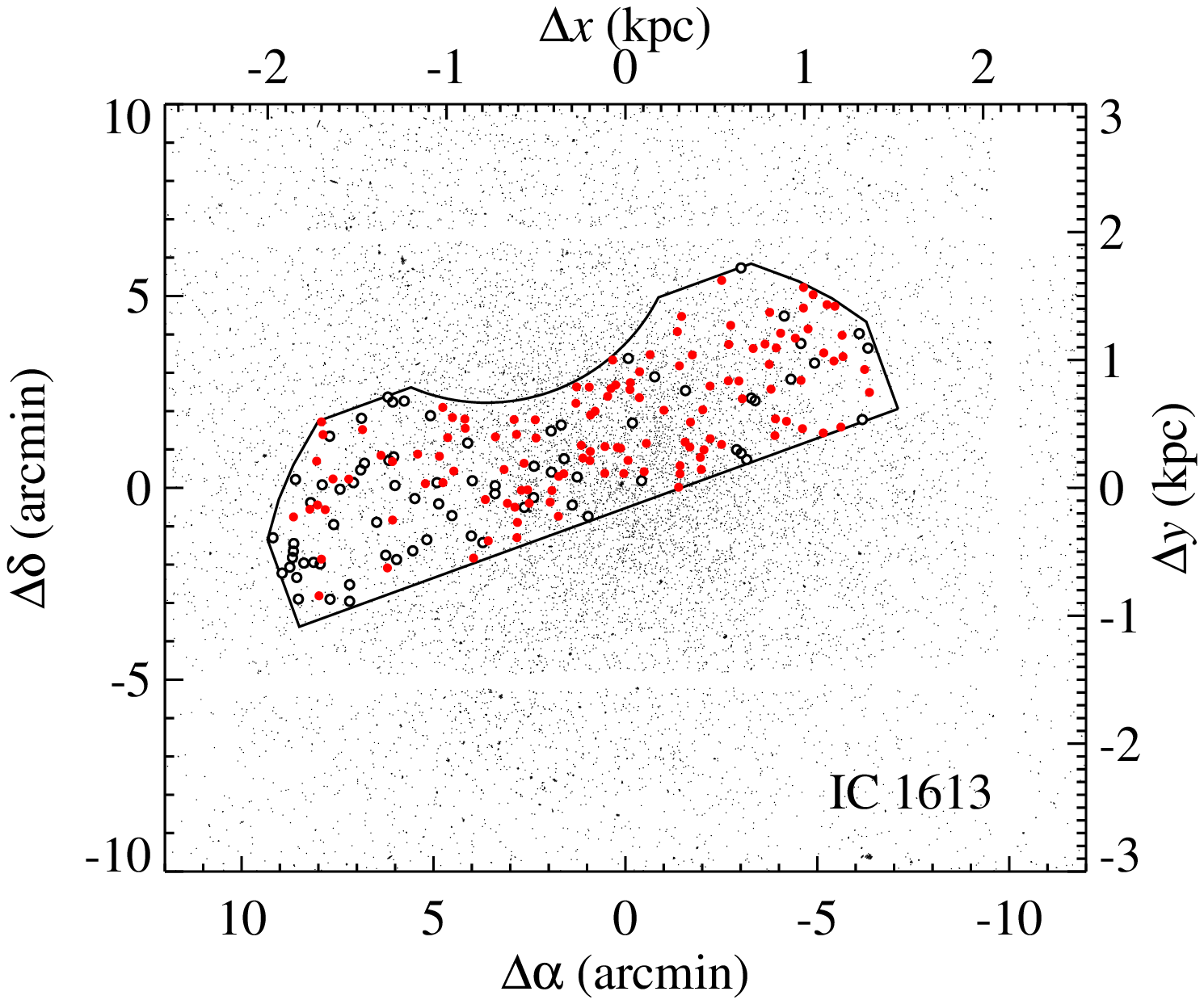}
\hfil
\includegraphics[width=0.46\textwidth]{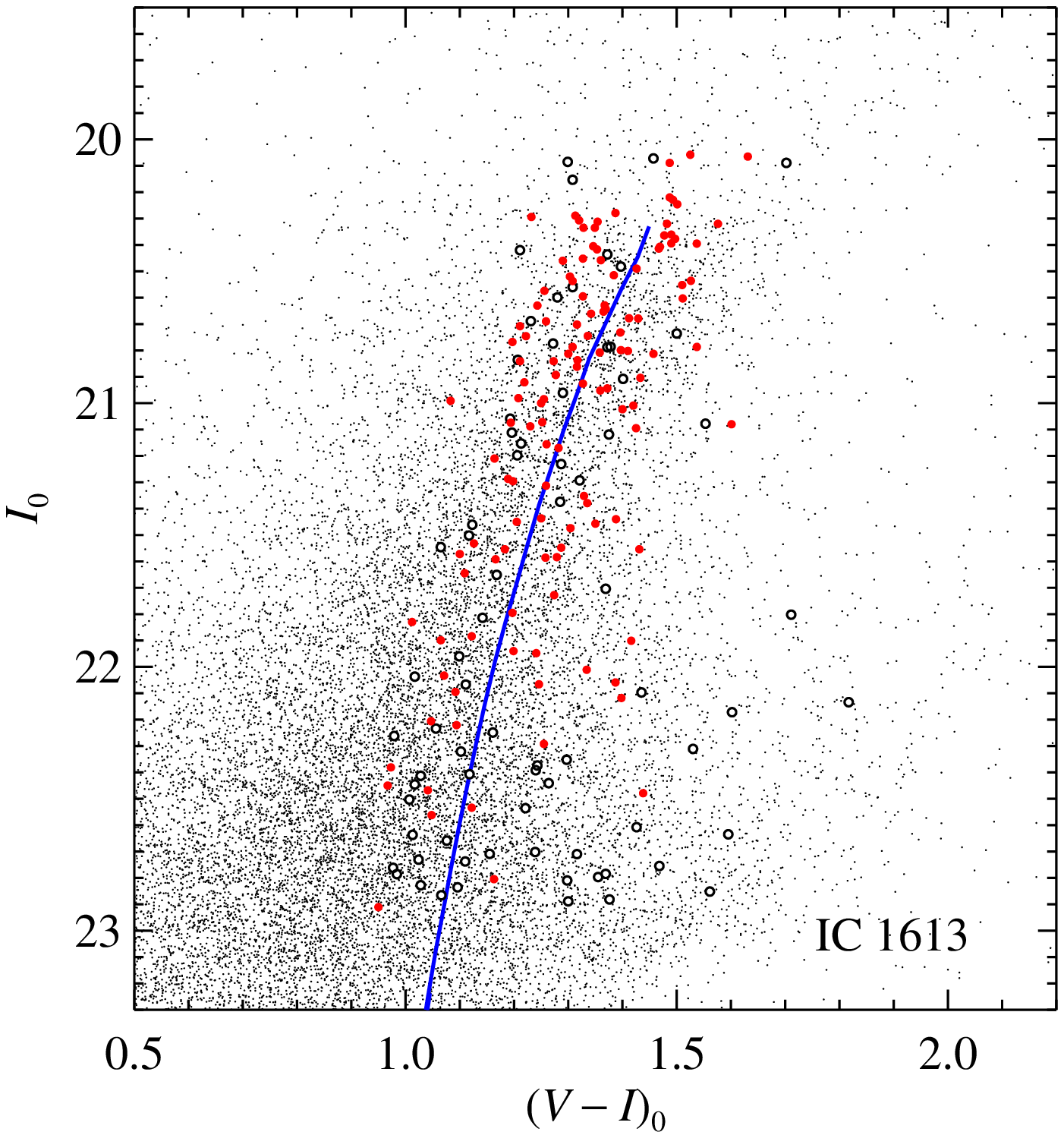}
\caption{Stars from the IC~1613 photometric catalog \protect
  \citep{ber07}.  The symbols are the same as in
  Figure~\ref{fig:n6822}.  The blue curve is a Padova theoretical
  isochrone \protect \citep{gir02} with an age of 4~Gyr and a
  metallicity of $\mathfeh = -1.2$.\label{fig:ic1613}}
\end{figure*}

\begin{figure*}[p]
\centering
\includegraphics[width=0.46\textwidth]{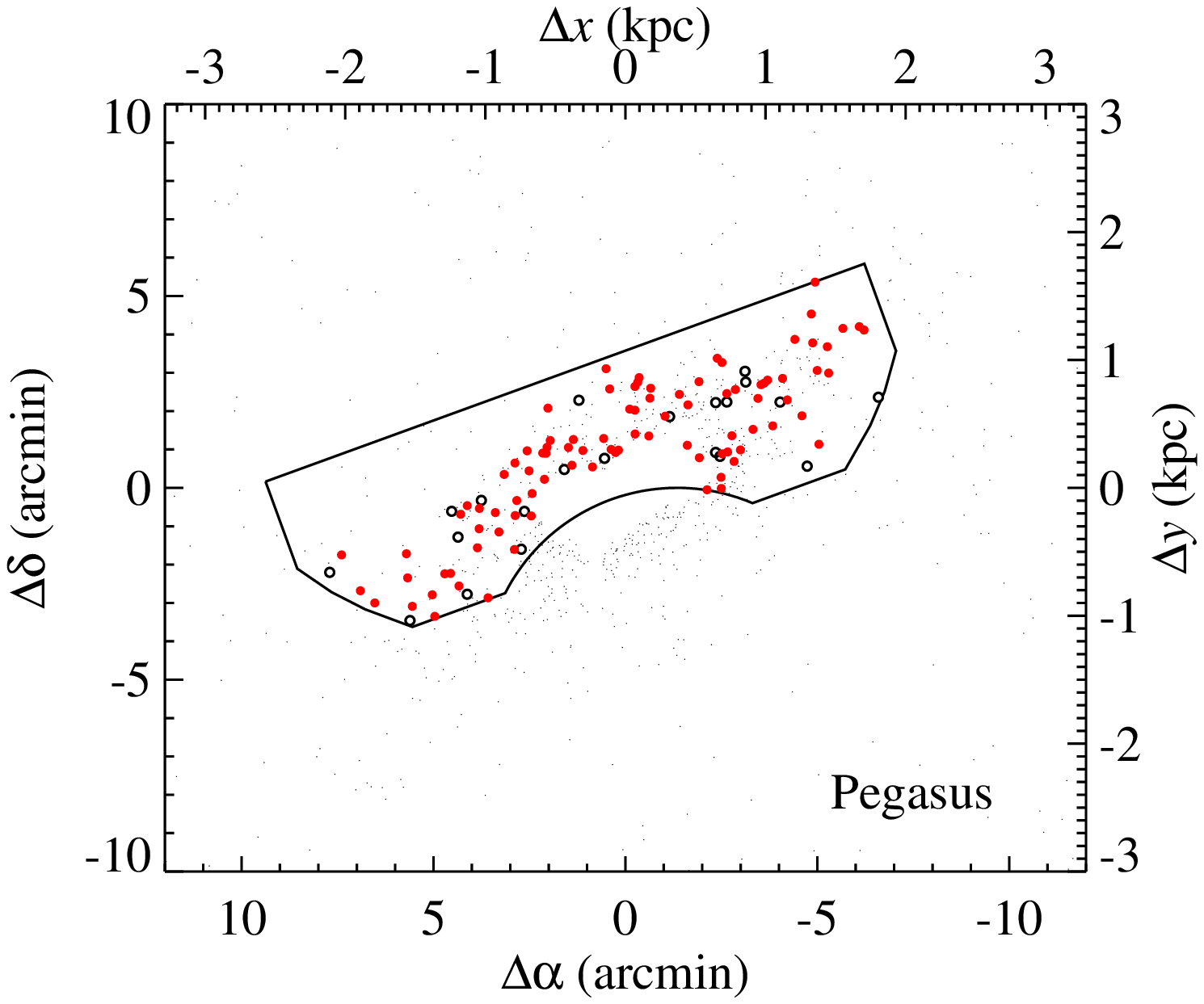}
\hfil
\includegraphics[width=0.46\textwidth]{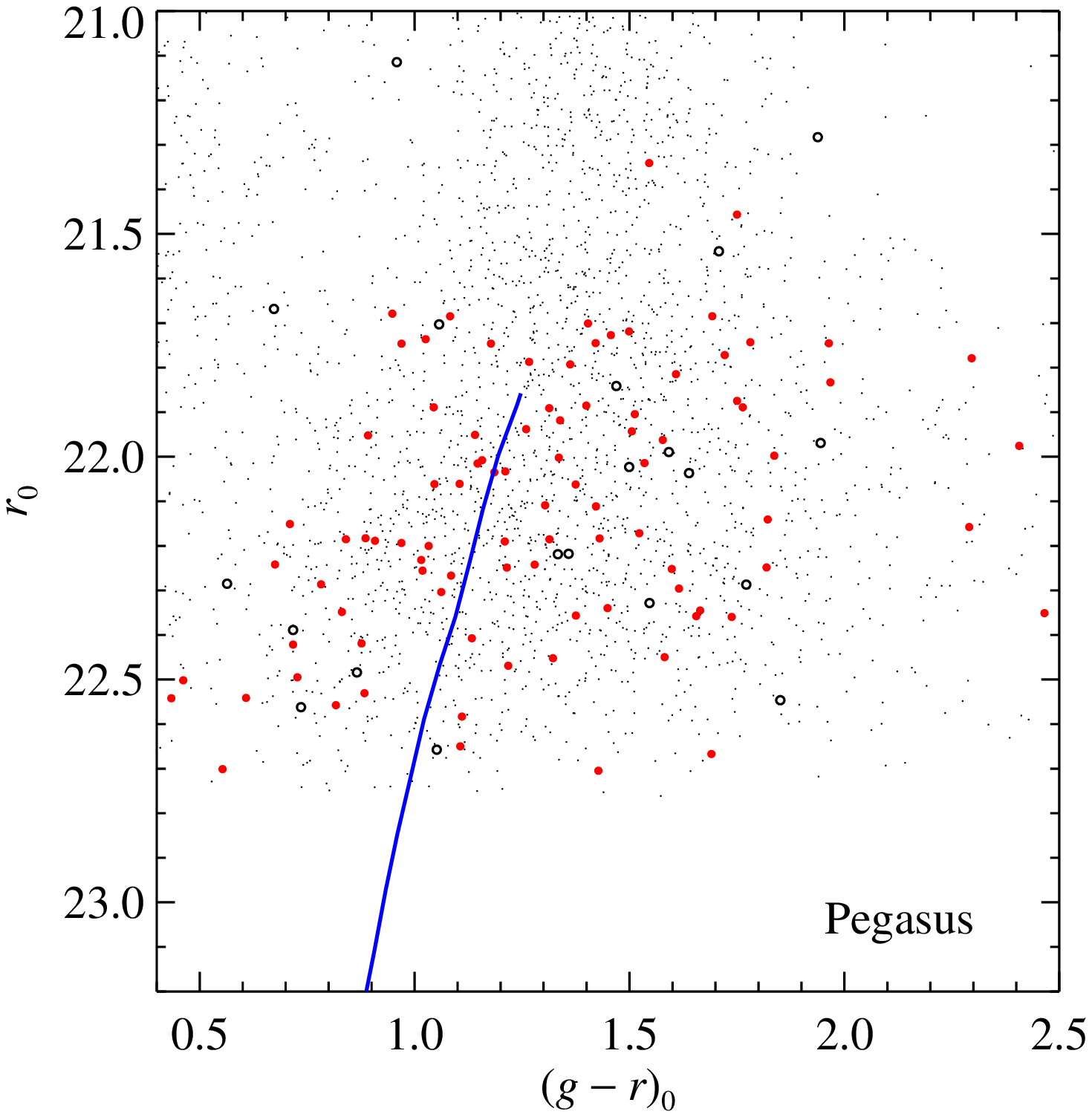}
\caption{Stars from the SDSS DR7 catalog \protect \citep{aba09}
  centered on the Pegasus dIrr.  Substantial photometric errors blur
  the shape of the RGB\@.  The symbols are the same as in
  Figure~\ref{fig:n6822}.  The blue curve is a Padova theoretical
  isochrone \protect \citep{gir04} with an age of 12~Gyr and a
  metallicity of $\mathfeh = -1.2$.\label{fig:peg}}
\end{figure*}

\begin{figure*}[p]
\centering
\includegraphics[width=0.46\textwidth]{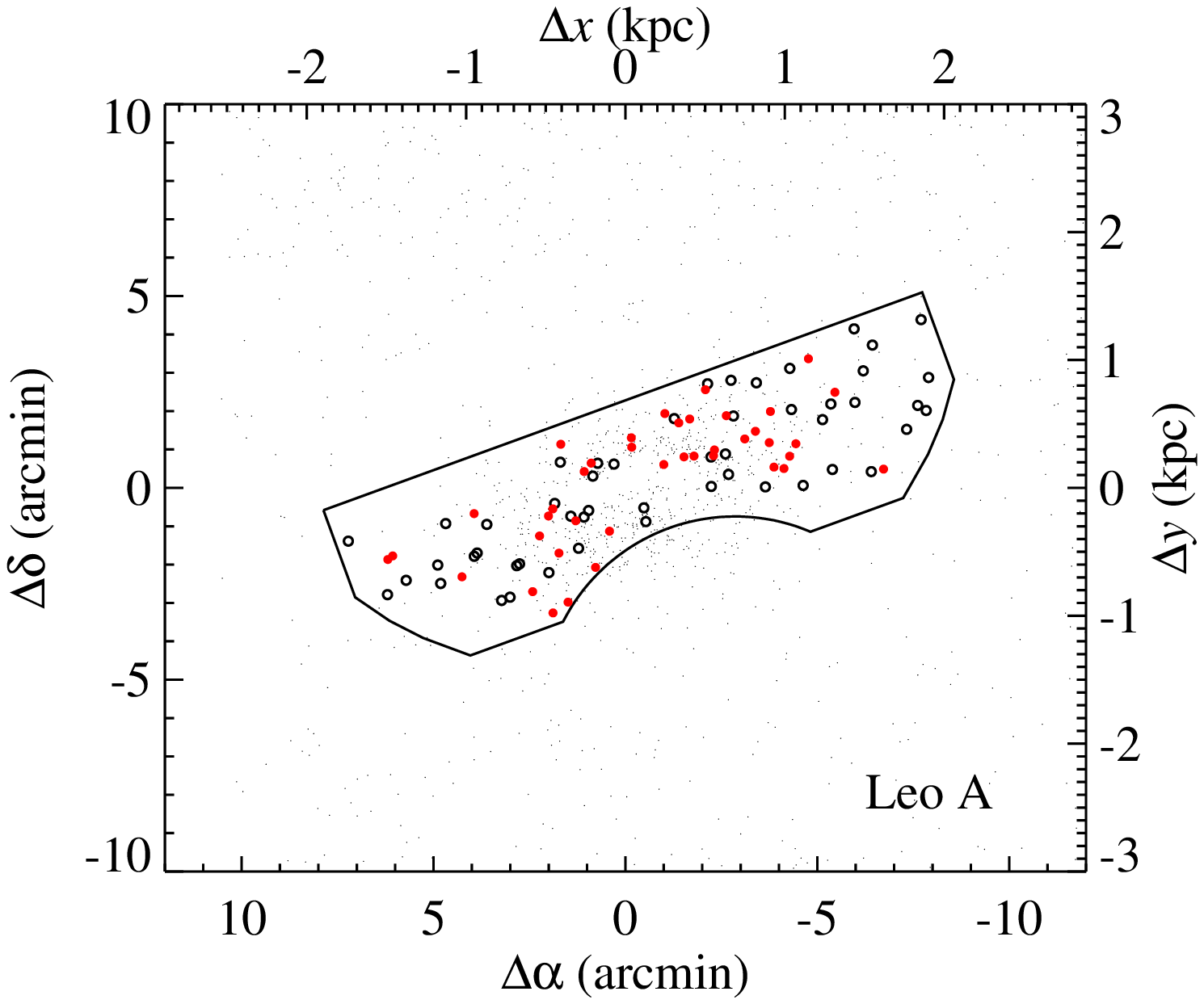}
\hfil
\includegraphics[width=0.46\textwidth]{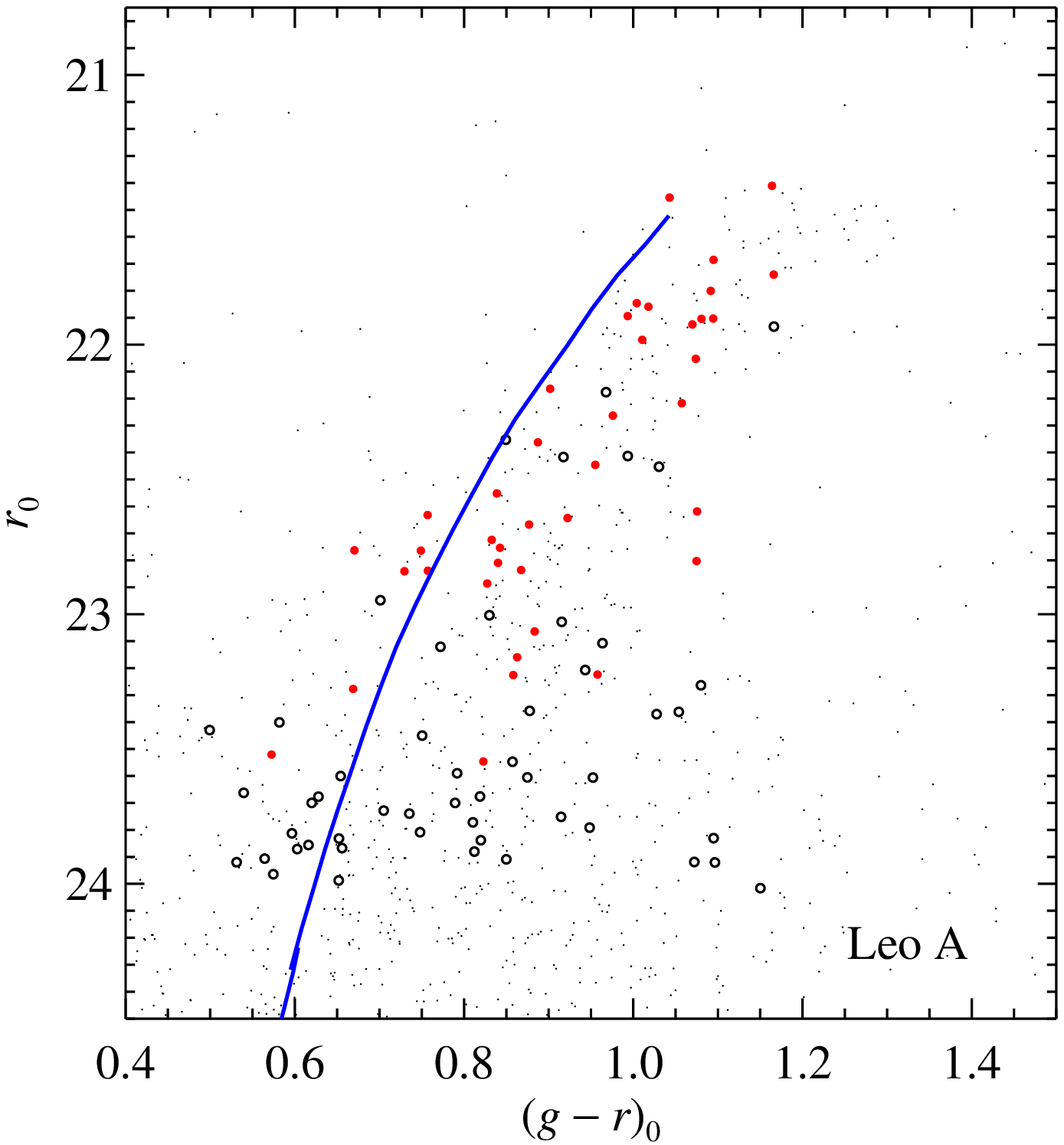}
\caption{Stars from the Leo~A photometric catalog \protect
  \citep{mcm01}.  The symbols are the same as in
  Figure~\ref{fig:n6822}.  The blue curve is a Padova theoretical
  isochrone \protect \citep{gir04} with an age of 8~Gyr and a
  metallicity of $\mathfeh = -1.6$.  Although the isochrone is bluer
  than most of the stars, an older, redder isochrone would be
  inconsistent with the photometrically measured SFH
  \citep{col07}.\label{fig:leoa}}
\end{figure*}

\begin{figure*}[ht!]
\centering
\includegraphics[width=0.46\textwidth]{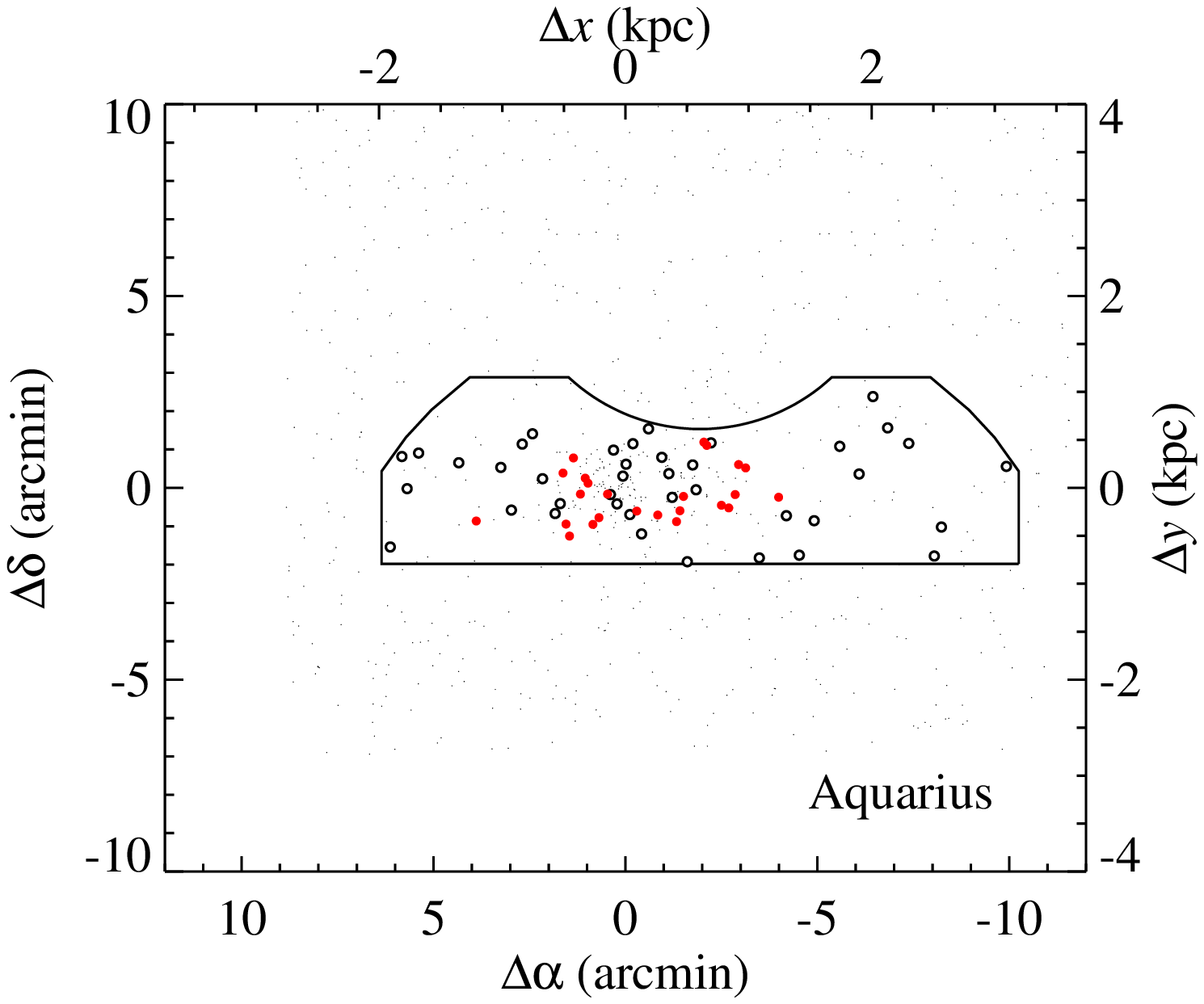}
\hfil
\includegraphics[width=0.46\textwidth]{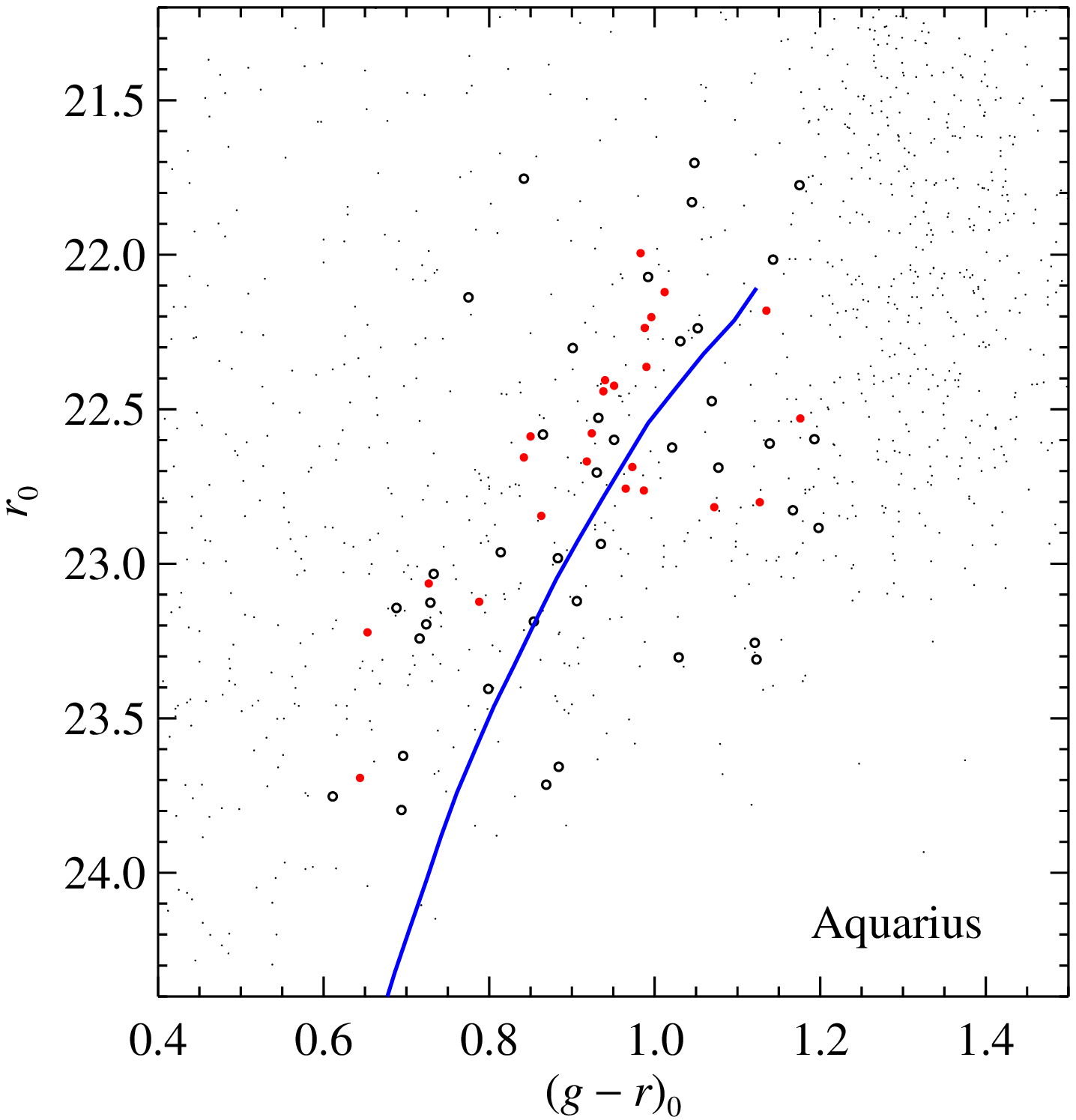}
\caption{Stars from the Aquarius photometric catalog.  The symbols are
  the same as in Figure~\ref{fig:n6822}.  The blue curve is a Padova
  theoretical isochrone \protect \citep{gir04} with an age of 2~Gyr
  and a metallicity of $\mathfeh = -1.2$.\label{fig:aqr}}
\end{figure*}

\begin{deluxetable*}{llccr@{ }c@{ }lcl}
\tablewidth{0pt}
\tablecolumns{9}
\tablecaption{DEIMOS Observations of dIrrs\label{tab:obs}}
\tablehead{\colhead{Galaxy} & \colhead{Slitmask} & \colhead{\# targets} & \colhead{Exp. Time} & \multicolumn{3}{c}{Date} & \colhead{Seeing} & \colhead{Individual Exposures} \\
\colhead{ } & \colhead{ } & \colhead{ } & \colhead{(h)} & \multicolumn{3}{c}{(UT)} & \colhead{($''$)} & \colhead{(s)}}
\startdata
NGC~6822    & n6822a  &    180 & \phn8.7 & 2010 & Aug & 12 & $1.2$ & $5 \times 1800 + 1630 + 1560 + 1200$ \\
            &         &        &         & 2011 & Jul & 30 & $0.7$ & $2 \times 1800$ \\
            &         &        &         & 2011 & Jul & 31 & $0.7$ & $5 \times 1800$ \\
            &         &        &         & 2011 & Aug & 6  & $1.0$ & $3 \times 1800$ \\
            & n6822b  &    180 & \phn6.0 & 2011 & Jul & 29 & $0.7$ & $8 \times 1800$ \\
            &         &        &         & 2011 & Aug & 4  & $0.9$ & $4 \times 1800$ \\
IC~1613     & i1613a  &    199 &    10.3 & 2010 & Aug & 11 & $0.9$ & $5 \times 1800 + 2 \times 1500$ \\
            &         &        &         & 2010 & Aug & 13 & $0.9$ & $3 \times 1800 + 720$ \\
            &         &        &         & 2011 & Jul & 29 & $0.7$ & $2 \times 1800 + 2 \times 1680$ \\
            &         &        &         & 2011 & Jul & 30 & $0.7$ & $3 \times 1800 + 1020$ \\
            &         &        &         & 2011 & Jul & 31 & $0.7$ & $3 \times 1800$ \\
VV~124\tablenotemark{a} & vv124a  &    121 & \phn3.7 & 2011 & Jan & 30 & $0.6$ & $6 \times 1800 + 2 \times 1200$ \\
            & vv124b  &    120 & \phn3.8 & 2011 & Jan & 30 & $0.6$ & $6 \times 1800 + 1600 + 1200$ \\
Pegasus~dIrr & pega    &    113 & \phn6.8 & 2011 & Aug & 4  & $0.7$ & $1800 + 1680 + 1380 + 600$ \\
            &         &        &         & 2011 & Aug & 5  & $0.9$ & $2 \times 1800 + 2 \times 1500 + 1200$ \\
            &         &        &         & 2011 & Aug & 6  & $1.0$ & $1800 + 1720 + 600$ \\
            &         &        &         & 2011 & Aug & 7  & $0.4$ & $1800 + 3 \times 1500 + 900$ \\
Leo~A       & leoaaW  & \phn91 & \phn6.7 & 2013 & Jan & 14 & $0.9$ & $12 \times 1800 + 2 \times 1200$ \\
Aquarius    & aqra    & \phn64 & \phn8.9 & 2013 & Jul & 8  & $0.5$ & $10 \times 1800 + 1260 + 720$ \\
            &         &        &         & 2013 & Sep & 1  & $0.7$ & $2 \times 1800$ \\
            &         &        &         & 2013 & Sep & 2  & $0.9$ & $3 \times 1800 + 2 \times 1560$ \\
Leo~T\tablenotemark{b} & LeoT--1 & \phn87 & \phn1.0 & 2007 & Feb & 14 & $0.7$ & $3 \times 1200$ \\
\enddata
\tablenotetext{a}{\citet{kir12} originally published these observations of VV124.}
\tablenotetext{b}{\citet{sim07} originally published these observations of Leo~T.}
\end{deluxetable*}

We observed dwarf galaxies spanning a range of luminosities and galaxy
types.  In order to maintain the homogeneity of the spectroscopic data
by using only Keck/DEIMOS \citep{fab03} spectroscopy, we observed only
galaxies visible from Mauna Kea.  Table~\ref{tab:sample} lists the
galaxies observed, separated by host (MW or M31) and galaxy type
(dSph/dE or dIrr).  The table includes the primary observables that
distinguish dSphs from dIrrs \citep[the June 2013 version of the
  compilation of][and references therein]{mcc12}.

DSphs and dEs are distinct from dIrrs in their morphology; distance
from a host galaxy, like the MW or M31; degree of rotation,
quantifiable as the ratio of the rotation speed to the velocity
dispersion ($v_{\rm rot}/\sigma_v$); and gas content.  The distinction
between dSphs and dEs is not clear.  The two classes have similar
surface brightness profiles, which are distinct from slightly larger
``true'' elliptical galaxies, like M32 \citep{wir84,kor09}.  The dEs
of M31 (NGC~147, 185, and 205) seem to be higher luminosity
counterparts to LG dSphs.  Therefore, we classify those three galaxies
as dSphs.

On the other hand, the dIrrs are not a homogeneous group.  Some dIrrs,
like VV~124, Pegasus, and Leo~T, share some properties with dSphs.
Their stellar and gas motions may be less dominated by rotation than
dispersion, and they may have lower gas fractions than typical dIrrs.
In accordance with the theory that dIrrs transform into dSphs by
interaction with a larger galaxy \citep{lin83,may01}, these galaxies
are called transition dwarf galaxies (dTs or dIrr/dSphs).
Alternatively, at least some dTs may simply be dIrrs that are
experiencing a temporary lull in SFR \citep{ski03}, but the
morphology--density relation indicates that proximity to a host plays
some role in making the transition from dSph to dIrr \citep{wei11}.
However, even dTs are starkly distinct from dSphs, especially in their
gas fractions.  All of the dSphs in our sample have $M_{\rm HI}/M_* <
0.002$ and all of the dIrrs and dTs have $M_{\rm HI}/M_* \ge 0.1$.
For simplicity, we conflate dIrrs and dTs into a single category
called dIrr.

Much of the spectroscopy and chemical analysis presented here has
already been published.  \citet{sim07} and \citet{kir10,kir13}
published the details of the observations and the analysis of the MW
dSphs.  \citeauthor{sim07}\ also included the Leo~T transition dwarf
galaxy in their sample.  \citet{kir12} described the VV~124 data and
analysis.  The M31 satellite sample comes from the Spectroscopic and
Panchromatic Landscape of Andromeda's Stellar Halo
\citep[SPLASH,][]{guh05,guh06}.  The details of the spectroscopy have
been published by \citet[][NGC~147, 185, and 205]{geh06,geh10},
\citet[][Andromeda~I, II, III, VII, and X]{kal09,kal10},
\citet[][additional Andromeda~II spectra]{ho12},
\citet[][Andromeda~XIV]{maj07}, and \citet[][Andromeda~V, IX, XV, and
  XVIII and additional spectroscopy of Andromeda~III and XIV]{tol12}.

We present new spectroscopy of red giants in NGC~6822, IC~1613, the
Pegasus dIrr, Leo~A, and Aquarius.  These are the first published
results from red giant spectroscopy in all of these galaxies except
NGC~6822, which was studied by \citet{tol01}.  The next section
describes our target selection for the new spectroscopy.

\subsection{Target Selection}
\label{sec:target}

We selected red giants for DEIMOS spectroscopy from existing
photometric catalogs of NGC~6822, IC~1613, Pegasus, and Leo~A\@.  We
obtained new images of Aquarius with Subaru/Suprime-Cam \citep{miy02}.

\subsubsection{NGC~6822}

We selected targets in NGC~6822 from \citeauthor{mas07}'s
(\citeyear{mas07}) $UBVRI$ photometric catalog.  The NGC~6822 data
came from the Mosaic camera on the Cerro Tololo Blanco 4~m telescope.
RGB candidates were selected from the CMD assuming a distance modulus
of 23.40, based on measurements of Cepheid variables \citep{fea12}.
No star with an apparent magnitude fainter than $I = 22$ was selected.
Magnitudes were corrected for extinction based on a reddening of
$E(B-V) = 0.25$ \citep{mas07}.  In order not to bias the sample with
respect to metallicity, color was not given much weight in the target
selection other than to select targets with the approximate colors of
red giants.  Figure~\ref{fig:n6822} shows the target selection in
celestial coordinates and in a CMD\@.  The figure also identifies
targets that passed the spectroscopic membership criteria
(Section~\ref{sec:membership}).  The theoretical isochrones shown in
Figure~\ref{fig:n6822} and in the subsequent figures are meant simply
to show the approximate shape of the RGB, not to indicate the true
ages of the galaxies.

\subsubsection{IC~1613}

We selected DEIMOS targets for IC~1613 from \citeauthor{ber07}'s
(\citeyear{ber07}) photometric catalog, kindly provided to us by
E.\ Bernard.  We assumed a Cepheid-based distance modulus of 24.34
\citep{tam11}.  The faint magnitude limit was $I = 23$.  The
extinction corrections were $A_V = 0.08$ and $A_I = 0.05$
\citep{sak04}.  Other details are the same as NGC~6822.
Figure~\ref{fig:ic1613} shows the target selection.

\subsubsection{Pegasus dIrr}

We consulted SDSS Data Release 7 \citep[DR7,][]{aba09} to generate a
photometric catalog for Pegasus.  We used CasJobs, the SDSS database
server, to download $ugriz$ magnitudes for all point sources
classified as stars within $30'$ of Pegasus.  The tip of the RGB in
Pegasus is close to SDSS's faint magnitude limit.  As a result, the
photometric errors are large, as revealed by the large scatter of
spectroscopic targets in color and magnitude (Figure~\ref{fig:peg}).
The Cepheid-based distance modulus is 24.87 \citep{tam11}.  We used
SDSS's extinction corrections, which are based on the \citet{sch98}
dust maps.  The faint magnitude limit, set by the SDSS photometric
depth, was $r = 23$.

\subsubsection{Leo~A}

We selected spectroscopic targets in Leo~A from the Isaac Newton
Telescope Wide Field Survey \citep{mcm01}.  A.\ Cole and M.\ Irwin
kindly provided the photometric catalog to us.  We imposed a faint
magnitude limit of $r = 24.1$.  We assumed the Cepheid-based distance
of 24.59 \citep{tam11}.  We corrected for extinction star by star
based on the \citet{sch98} dust maps.  Other details are the same as
NGC~6822.  Figure~\ref{fig:leoa} shows the targets.

\subsubsection{Aquarius}
\label{sec:aqr}

We observed Aquarius with Suprime-Cam on 2010 May 25 in $0.8''$
seeing.  We obtained five 60~s exposures in each of the $g$ and $r$
filters for a total of five minutes in each filter.  The images were
reduced with the \mbox{SDFRED2} software \citep{ouc04}.  We identified
point sources and calculated photometric magnitudes using DAOPHOT
\citep{ste87} within version 2.12.2 of IRAF \citep{tod86}.

We selected targets for spectroscopy in a manner similar to the other
dIrrs.  The assumed distance modulus, based on the magnitude of the
tip of the RGB, was 25.15 \citep{mcc06}.  We imposed a faint magnitude
limit of $r = 24$.  We corrected magnitudes for extinctions of $A_g =
0.20$ and $A_r = 0.13$ \citep{sch98}.  Figure~\ref{fig:aqr} shows the
astrometry and photometry for spectroscopic targets.

\subsection{DEIMOS Spectroscopy}

We observed the new dIrr slitmasks with DEIMOS in the summers of 2010,
2011, and 2013.  We observed two slitmasks for NGC~6822 and VV~124 and
one slitmask each for the remaining dIrrs.  Table~\ref{tab:obs} gives
the number of red giant candidates, the exposure time, the observation
date, and the seeing for each slitmask.

We configured the spectrograph in the same way for all of the
slitmasks.  We used the 1200G grating, which has a groove spacing of
1200~mm$^{-1}$ and a blaze wavelength of 7760~\AA\@.  The grating was
tilted to a central wavelength of 7800~\AA, which resulted in a
spectral range of roughly 6400--9000~\AA\@.  The exact spectral range
for each object depended on the placement of the slit on the slitmask.
The slit width was $0.7''$ for all slitmasks except leoaaW.  For
Leo~A, we used a backup slitmask with $1.1''$ slits because the seeing
exceeded $1''$ when we started observing.  The resolution of DEIMOS in
this configuration yields a line profile of 1.2~\AA\ FWHM for the
$0.7''$ slits and 1.8~\AA\ FWHM for the $1.1''$ slits.  These
resolutions correspond to resolving powers of $R \sim 7100$ and 4700,
respectively, at 8500~\AA, near the CaT\@.  We reduced the data into
sky-subtracted, one-dimensional spectra with the spec2d software
pipeline \citep{coo12,new13}.  For slightly more details on the
observing and reduction procedures, see \citet{kir12}.

%%%%%%%%%%%%%%%%%%%%%%%%%%%%%%%%%
%%%%%%%%%   SECTION 3   %%%%%%%%%
%%%%%%%%%%%%%%%%%%%%%%%%%%%%%%%%%

\section{Metallicity Measurements}
\label{sec:spectroscopy}

\addtocounter{table}{1}

\subsection{Membership}
\label{sec:membership}

We removed contaminants that do not belong to the galaxies from the
spectroscopic sample in order to have clean metallicity distributions.
We imposed membership cuts based on the CMD, spectral features, and
radial velocities.

The CMD membership cut was very lax.  Stars that could plausibly be
members of the RGB in each galaxy were allowed.  No culling based on
the CMD was performed after the slitmasks were designed.  This
liberalism with color selection minimizes selection bias in the
metallicity distribution (but see Section~\ref{sec:bias}).  The
selection of stars based on colors and magnitudes is shown in
Figures~\ref{fig:n6822} through \ref{fig:aqr}.

The membership cut based on spectral features was also not stringent.
We discarded a few stars with very strong \ion{Na}{1}~8190 doublets,
which happen only in dwarf stars with high surface gravities
\citep{spi71}.  All of these stars would also have been ruled
non-members on the basis of radial velocity.

The primary membership cut was radial velocity.  We used the same
procedure for determining velocity membership as \citet{kir10}.  The
radial velocities were measured by cross-correlating the observed
spectra with DEIMOS template spectra in the wavelength range of the
CaT\@.  We used the same templates as \citet{sim07}.  For each galaxy,
we started with a guess at the average velocity ($v_0$) and velocity
dispersion ($\sigma_v$).  Then, we discarded all stars more than
$3\sigma_v$ discrepant from $v_0$.  We recalculated $v_0$ and
$\sigma_v$ from the culled sample, and we repeated the process until
it converged.  The stars in each galaxy that passed this iterative
membership cut comprise the final member samples.

\subsection{Individual Stellar Metallicities}
\label{sec:individualmetallicities}

We measured spectroscopic metallicities for individual stars in the
dIrrs and the MW satellites.  \citet{kir08b,kir09,kir10,kir12,kir13}
published the metallicities for the MW satellites, VV~124, and
Leo~T\@.  Here, we present new measurements for the remaining dIrrs.

\citet{kir08a,kir09,kir10} detailed the technique for measuring
metallicities for individual stars.  It is based on spectral synthesis
of \ion{Fe}{1} absorption lines.  First, the continuum of the observed
spectrum was shifted to the rest frame and normalized to unity.  Then,
it was compared to a large grid of synthetic spectra.  The search for
the best-fitting synthetic spectrum was a minimization of $\chi^2$ of
pixels around \ion{Fe}{1} lines.

Photometry helped to constrain the surface gravity and effective
temperature.  The surface gravity was fixed with the help of 14~Gyr
theoretical isochrones and the observed color and magnitude of the
star.  The same method was used to determine a first guess at the
effective temperature, but the temperature was allowed to vary during
the spectral fitting within a range around the photometric
temperature.  The amount by which the spectroscopic temperature was
allowed to stray from the photometric temperature depended on the
magnitude of the error in the photometric color.  For galaxies with
photometry in the Johnson/Cousins filter set, we used Yonsei-Yale
isochrones \citep{dem04}.  For photometry in the SDSS filter set, we
used Padova isochrones \citep{gir04}, which were computed with SDSS
filter transmission curves.  This procedure is slightly different from
our earlier metallicity catalogs \citep{kir08b,kir10}, where we
transformed SDSS magnitudes to Johnson/Cousins magnitudes.  As a
result, some of the average metallicities for galaxies are slightly
different, especially for the ultra-faint galaxies \citep{kir08b}.  We
made this revision to eliminate any potential errors caused by the
transformation of colors.

After the best-fitting temperature and metallicity were found, the
[$\alpha$/Fe] ratio was measured from neutral Mg, Si, Ca, and Ti
lines.  Next, the metallicity measurement was refined based on the
measured [$\alpha$/Fe] ratio.  The process was repeated until neither
[Fe/H] nor [$\alpha$/Fe] changed between iterations.  Finally,
[Mg/Fe], [Si/Fe], [Ca/Fe], and [Ti/Fe] were measured individually.
This paper presents measurements of [Fe/H] only.  The [$\alpha$/Fe]
ratios will be used in a different study of the SFHs of the dIrrs,
similar to \citeauthor{kir11b}'s (\citeyear{kir11b}) study of the SFHs
of the MW dSph satellites.

We discovered a minor error in the metallicity measurements published
by \citet{kir10}.  The $[\alpha/{\rm Fe}]_{\rm atm}$ ratio sometimes
fixed itself on spurious $\chi^2$ minima.  This error affected the
measurement of [Fe/H] by a small amount, on the order of 0.1--0.2~dex.
This study includes the corrected measurements.

Table~\ref{tab:catalog} gives the coordinates, extinction-corrected
magnitudes, temperatures, gravities, microturbulent velocities, and
metallicities for all of the individual member stars in the dIrrs and
the MW satellites.  The photometric filter set varies from galaxy to
galaxy.  Table~\ref{tab:catalog} includes Washington $M$ and $T_2$
magnitudes; Johnson/Cousins $B$, $V$, $R$, and $I$ magnitudes; and
SDSS $g$, $r$, and $i$ magnitudes.  The metallicities for the MW
satellites have been corrected for the aforementioned error.

\subsection{Coadded Stellar Metallicities}

\begin{figure}[t!]
\centering
\includegraphics[width=\columnwidth]{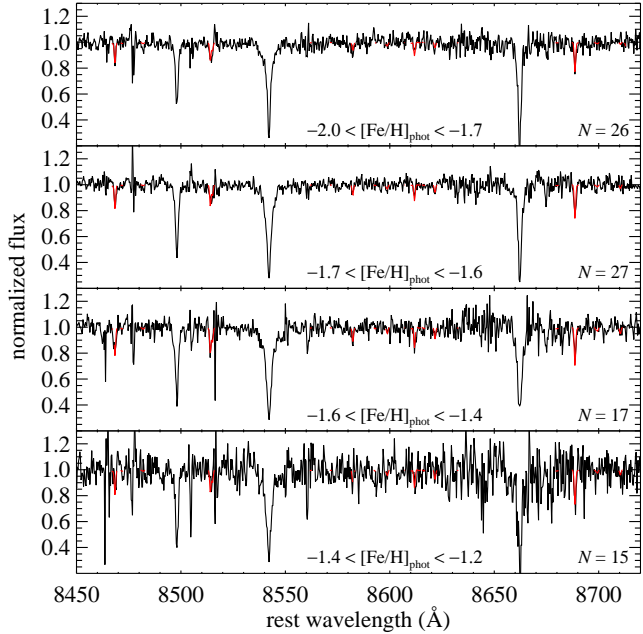}
\caption{Small portions of coadded spectra (black) for the four bins
  of stars in Andromeda~V\@.  Portions of the best-fitting, coadded,
  synthetic spectra that were used for determining [Fe/H] from
  \ion{Fe}{1} lines are shown in red.  Each panel gives the
  photometric metallicity range of the bin and the number of stars in
  the bin.\label{fig:coadd}}
\end{figure}

The M31 satellite spectra are too noisy to permit metallicity
measurements of individual stars.  We coadded the spectra in order to
attain signal-to-noise ratios (S/Ns) sufficient for comparing to the
grid of synthetic spectra.  \citet{yan13} described this technique and
demonstrated that it is effective at recovering the average
metallicities for groups of stars.  We applied their procedure to the
M31 satellite spectra.

The stars were selected for membership based on position in the CMD
and radial velocity \citep[see][]{tol12}.  The measurement of
velocities does not require S/N as high as the measurement of
metallicity.  Hence, velocities can be determined for individual
stars.  These velocities and the associated membership cuts were given
by the references listed in Section~\ref{sec:galaxies}.

The binning was based on photometric metallicity ($\mathfeh_{\rm
  phot}$), which was determined by comparing the de-reddened color and
extinction-corrected magnitude of each star to a set of theoretical
Yonsei--Yale isochrones \citep{dem04} with an age of 14~Gyr.  This
choice of age affects the binning but not the spectroscopic
metallicity.  By interpolating in the CMD between isochrones of
different metallicities, a value of $\mathfeh_{\rm phot}$ can be
assigned to each star.  The stars were binned in $\mathfeh_{\rm
  phot}$.  This procedure is similar to binning by color, but it
accounts for the curvature of the RGB in the CMD\@.  The bins were
chosen to have a minimum width of $\Delta \mathfeh_{\rm phot} = 0.1$
and a minimum of 15 stars per bin.  Table~\ref{tab:lzr} lists the
number of bins in each M31 satellite as well as the total number of
stars across all bins.

Because the spectral shape varies from star to star, continuum
normalization needed to be treated carefully.  The continua of the
individual spectra were determined by fitting a B-spline to the
spectra, masking out regions of telluric absorption and stellar
absorption lines \citep[see][]{kir10}.  The spectra were divided by
their continua and rebinned onto a common rest wavelength array.  They
were stacked with inverse variance weighting on each pixel.  The
stacked spectrum is called $s_1$.

To refine the continuum determination, each individual, un-normalized
spectrum was divided by $s_1$.  B-splines were fit to the residual
spectra, but only the strongest absorption lines (those reaching a
15\% flux decrement in $s_1$) were masked.  These splines served as
the new continua for the individual spectra.  The new
continuum-divided individual spectra were stacked using a median
rather than a weighted mean.  The median spectrum is called $s_2$.

The coaddition was refined a third time to remove noise spikes from
improperly subtracted sky lines, cosmic rays, and other artifacts.
The individual spectra were continuum-divided as in the previous
round, but pixels in the individual spectra more than $5\sigma$
discrepant from $s_2$ were masked.  The final coadded spectrum, $s_3$,
is a median stack of the individual spectra with $5\sigma$ clipping.
Figure~\ref{fig:coadd} shows an example of the four coadded spectra in
the M31 satellite Andromeda~V\@.

The rest of our procedure followed \citet{yan13}.  The coadded
spectrum, $s_3$, was compared to coadded synthetic spectra from the
spectral grid described in Section~\ref{sec:individualmetallicities}.
Each coadded synthetic spectrum was composed of the same number of
spectra as the coadded observed spectrum.  The effective temperature
and surface gravity of each star in the synthetic coaddition matched
the photometric temperature and gravity of each corresponding star in
the observed coaddition.  The metallicity was the same for each star
in the synthetic coaddition.  This metallicity was varied to minimize
$\chi^2$.

The metallicities of the coadded spectra show no systematic difference
with the central photometric metallicity of the bin.  The rms between
the spectral and photometric metallicities is about 0.2~dex for any
particular M31 satellite.  Even so, the coadded, spectroscopic
metallicities should be regarded as more accurate because they do not
suffer from the age--metallicity degeneracy.  Although the stars were
grouped according to $\mathfeh_{\rm phot}$, that value does not enter
into the determination of the spectroscopic metallicity.  The
metallicity determination relies on the photometric temperatures and
gravities of the stars, but those values are much less sensitive to
age than $\mathfeh_{\rm phot}$ is.

\subsection{Selection Biases}
\label{sec:bias}

Certain aspects of our experimental design could bias our metallicity
distributions against metal-rich or metal-poor stars.  Although we
performed almost no selection on color within the reasonable bounds of
the RGB, the photometric catalogs have a color bias for the faintest
stars.  For example, the photometry for NGC~6822 has a magnitude limit
of $V_0 \sim 22.3$ and $I_0 \sim 22.5$.  Stars fainter than these
limits have photometric errors greater than 0.1~dex.  The dashed, gray
line in Figure~\ref{fig:n6822} shows the $V_0$ limit, which is more
restrictive than the $I_0$ limit for RGB colors.  The magnitude limit
imposes a slight color bias against red stars for the faintest stars
in our NGC~6822 sample.  This bias is very small because many stars
fainter than the 0.1~dex error limit are still included in our sample.
Furthermore, the excluded portion of the RGB---had we been able to
include it---would have comprised less than 5\% of our spectroscopic
sample.  We deem this bias as negligible.

There is also a bias in the sense that galaxies have radial
metallicity gradients \citep[e.g.,][]{meh03,kir11a}.  When a galaxy
has a metallicity gradient, it is almost always in the sense that the
innermost stars are more metal-rich than the outermost stars.  Our
sample could have a bias against the outer, metal-poor stars because
most of our slitmasks were placed at the centers of their respective
galaxies.  This bias is especially applicable to the dSphs, which are
closer than the dIrrs and therefore have larger angular sizes.
Fortunately, most of the stars in both dIrrs and dSphs are
concentrated toward their centers.  Our slitmasks span at least the
half-light radii for all galaxies in our sample except Sextans.
Therefore, our samples represent at least half of the stellar
populations---and typically much more than half---in all of the
galaxies except for Sextans, which has a radial metallicity gradient
of 0.35~dex~kpc$^{-1}$ \citep{bat11}.  If the gradients that
\citet{kir11a} and \citet{bat11} measured persist out the tidal radii,
then the typical bias in $\langle \mathfeh \rangle$ caused by the
central concentration of our spectroscopic sample is $\la 0.15$~dex.

Some galaxies also show a correlation between metallicity and velocity
dispersion of distinct kinematic populations
\citep{tol04,bat06,bat11,wal11}.  Usually, the more metal-poor
population is dynamically hotter.  Applying a membership cut in
velocity space could bias the sample against metal-poor stars.
However, our $3\sigma_v$ velocity cut is quite inclusive.  If the
velocities are normally distributed among a single population, then
our velocity criterion includes 99.7\% of member stars.  Although
Fornax, Sculptor, and Sextans do not have single kinematic
populations, our velocity criterion includes 98.4\%, 99.0\%, and
96.9\%, respectively, of possible members within $5\sigma_v$ of $v_0$.
The Sextans spectroscopic sample is expected to be more contaminated
with non-members than the other two dSphs because it has the lowest
galactic latitude.  Even if all of the discarded stars belonged to
Sextans, the bias against metal-poor stars is at most 3.1\%\@.

We conclude that selection biases do not significantly affect the
average metallicities discussed in Section~\ref{sec:mzr}.  However,
selection biases could subtly affect the metallicity distributions
(Section~\ref{sec:gce}).  Our sample might be missing some of the
rare, extremely metal-poor stars that preferentially inhabit the dwarf
galaxies' outskirts.  It is also these stars that would be lost first
in tidal stripping as the dSphs fell into the MW\@.  This bias is
difficult to quantify because it depends on the dSph orbits and the
shape of the radial metallicity gradient out to the tidal radii, both
of which are unknown or poorly known for most dwarf galaxies.
Nonetheless, this bias affects only the detailed shape of the
metal-poor part of the metallicity distributions, not average
metallicities or the bulk of the metallicity distributions.

%%%%%%%%%%%%%%%%%%%%%%%%%%%%%%%%%
%%%%%%%%%   SECTION 4   %%%%%%%%%
%%%%%%%%%%%%%%%%%%%%%%%%%%%%%%%%%

\section{Mass--Metallicity Relation}
\label{sec:mzr}

\tabletypesize{\scriptsize}

\begin{deluxetable*}{lcccccccccc}
\tablecolumns{11}
\tablewidth{0pt}
\tablecaption{Summary of MDFs\label{tab:lzr}}
\tablehead{\colhead{Galaxy} & \colhead{$N$\tablenotemark{a}} & \colhead{$\log (L_V/L_{\sun})$} & \colhead{$\log (M_*/M_{\sun})$} & \colhead{$\langle {\rm [Fe/H]}\rangle$\tablenotemark{b}} & \colhead{$\sigma$\tablenotemark{c}} & \colhead{Median} & \colhead{mad\tablenotemark{d}} & \colhead{IQR\tablenotemark{e}} & \colhead{Skewness} & \colhead{Kurtosis\tablenotemark{f}}}
\startdata
\cutinhead{Milky Way dSphs}
Fornax            &         672 &        $7.31 \pm 0.14$ &        $7.39 \pm 0.14$                  & $-1.04 \pm 0.01$ & $0.33$ $(0.29)$ & $-1.06$ & $0.17$ & $0.34$ &     $-1.29 \pm 0.09$ & \phs$ 3.80 \pm 0.19$ \\
Leo~I             &         814 &        $6.74 \pm 0.13$ &        $6.69 \pm 0.13$                  & $-1.45 \pm 0.01$ & $0.32$ $(0.28)$ & $-1.44$ & $0.18$ & $0.36$ &     $-1.43 \pm 0.09$ & \phs$ 4.66 \pm 0.17$ \\
Sculptor          &         375 &        $6.36 \pm 0.21$ &        $6.59 \pm 0.21$                  & $-1.68 \pm 0.01$ & $0.46$ $(0.44)$ & $-1.65$ & $0.33$ & $0.71$ &     $-0.70 \pm 0.13$ & \phs$ 0.36 \pm 0.25$ \\
Leo~II            &         256 &        $5.87 \pm 0.13$ &        $6.07 \pm 0.13$                  & $-1.63 \pm 0.01$ & $0.40$ $(0.36)$ & $-1.61$ & $0.22$ & $0.47$ &     $-1.15 \pm 0.15$ & \phs$ 1.40 \pm 0.30$ \\
Sextans           &         123 &        $5.64 \pm 0.20$ &        $5.84 \pm 0.20$                  & $-1.94 \pm 0.01$ & $0.47$ $(0.38)$ & $-1.96$ & $0.26$ & $0.56$ &     $-0.12 \pm 0.22$ & \phs$ 0.44 \pm 0.43$ \\
Ursa Minor        &         190 &        $5.45 \pm 0.20$ &        $5.73 \pm 0.20$                  & $-2.13 \pm 0.01$ & $0.43$ $(0.34)$ & $-2.12$ & $0.24$ & $0.48$ & \phs$ 0.60 \pm 0.18$ & \phs$ 2.52 \pm 0.35$ \\
Draco             &         269 &        $5.43 \pm 0.10$ &        $5.51 \pm 0.10$                  & $-1.98 \pm 0.01$ & $0.42$ $(0.35)$ & $-1.97$ & $0.24$ & $0.48$ &     $-0.35 \pm 0.15$ & \phs$ 0.43 \pm 0.30$ \\
Canes Venatici~I  &         151 &        $5.36 \pm 0.09$ &        $5.48 \pm 0.09$                  & $-1.91 \pm 0.01$ & $0.44$ $(0.39)$ & $-1.88$ & $0.29$ & $0.56$ &     $-0.12 \pm 0.20$ & \phs$ 0.26 \pm 0.39$ \\
Hercules          &     \phn 19 &        $4.56 \pm 0.14$ &        $4.57 \pm 0.14$                  & $-2.39 \pm 0.04$ & $0.51$ $(0.45)$ & $-2.48$ & $0.41$ & $0.83$ & \phs$ 0.64 \pm 0.52$ &     $-0.66 \pm 1.01$ \\
Ursa Major~I      &     \phn 28 &        $4.15 \pm 0.13$ &        $4.28 \pm 0.13$                  & $-2.10 \pm 0.03$ & $0.65$ $(0.60)$ & $-2.39$ & $0.52$ & $1.01$ & \phs$ 0.40 \pm 0.44$ &     $-0.72 \pm 0.86$ \\
Leo~IV            & \phn\phn  9 &        $3.94 \pm 0.16$ & $3.93^{+0.15}_{-0.11}$                  & $-2.45 \pm 0.07$ & $0.65$ $(0.59)$ & $-2.37$ & $0.49$ & $0.94$ & \phs$ 0.47 \pm 0.72$ &     $-1.26 \pm 1.40$ \\
Canes Venatici~II &     \phn 14 &        $3.90 \pm 0.20$ &        $3.90 \pm 0.20$                  & $-2.12 \pm 0.05$ & $0.59$ $(0.57)$ & $-2.39$ & $0.36$ & $0.71$ & \phs$ 0.66 \pm 0.60$ &     $-0.64 \pm 1.15$ \\
Ursa Major~II     &     \phn 11 &        $3.60 \pm 0.23$ &        $3.73 \pm 0.23$                  & $-2.18 \pm 0.05$ & $0.66$ $(0.60)$ & $-2.30$ & $0.44$ & $0.60$ & \phs$ 0.53 \pm 0.66$ &     $-0.93 \pm 1.28$ \\
Coma Berenices    &     \phn 19 &        $3.57 \pm 0.22$ &        $3.68 \pm 0.22$                  & $-2.25 \pm 0.04$ & $0.43$ $(0.39)$ & $-2.44$ & $0.30$ & $0.52$ & \phs$ 0.27 \pm 0.52$ &     $-0.59 \pm 1.01$ \\
Segue~2           & \phn\phn  8 &        $2.93 \pm 0.13$ &        $3.14 \pm 0.13$\tablenotemark{g} & $-2.14 \pm 0.05$ & $0.38$ $(0.33)$ & $-2.20$ & $0.33$ & $0.62$ & \phs$ 0.51 \pm 0.75$ &     $-1.16 \pm 1.50$ \\
\cutinhead{Local Group dIrrs}
NGC~6822          &         278 &        $8.02 \pm 0.09$ &        $7.92 \pm 0.09$                  & $-1.05 \pm 0.01$ & $0.49$ $(0.47)$ & $-1.02$ & $0.28$ & $0.60$ &     $-0.84 \pm 0.15$ & \phs$ 1.38 \pm 0.29$ \\
IC~1613           &         125 &        $8.01 \pm 0.06$ &        $8.01 \pm 0.06$                  & $-1.19 \pm 0.01$ & $0.37$ $(0.32)$ & $-1.22$ & $0.23$ & $0.47$ &     $-0.36 \pm 0.22$ &     $-0.09 \pm 0.43$ \\
VV~124            &     \phn 52 &        $6.92 \pm 0.08$ &        $6.92 \pm 0.08$\tablenotemark{h} & $-1.43 \pm 0.02$ & $0.52$ $(0.55)$ & $-1.53$ & $0.32$ & $0.68$ &     $-0.53 \pm 0.33$ &     $-0.26 \pm 0.65$ \\
Pegasus~dIrr      &     \phn 95 &        $6.82 \pm 0.08$ &        $6.82 \pm 0.08$                  & $-1.39 \pm 0.01$ & $0.56$ $(0.54)$ & $-1.31$ & $0.33$ & $0.68$ &     $-1.04 \pm 0.25$ & \phs$ 0.85 \pm 0.49$ \\
Leo~A             &     \phn 39 &        $6.78 \pm 0.09$ &        $6.47 \pm 0.09$                  & $-1.58 \pm 0.02$ & $0.42$ $(0.36)$ & $-1.67$ & $0.21$ & $0.42$ &     $-0.25 \pm 0.38$ &     $-0.39 \pm 0.74$ \\
Aquarius          &     \phn 24 &        $6.19 \pm 0.05$ &        $6.15 \pm 0.05$                  & $-1.44 \pm 0.03$ & $0.35$ $(0.31)$ & $-1.47$ & $0.28$ & $0.51$ & \phs$ 0.20 \pm 0.47$ &     $-0.95 \pm 0.92$ \\
Leo~T             &     \phn 16 &        $5.13 \pm 0.20$ &        $5.13 \pm 0.20$\tablenotemark{h} & $-1.74 \pm 0.04$ & $0.54$ $(0.47)$ & $-1.76$ & $0.16$ & $0.70$ &     $-0.68 \pm 0.56$ &     $-0.71 \pm 1.09$ \\
\cutinhead{M31 dSphs from Coadded Spectra}
NGC~205           &     11 /     334 &        $8.52 \pm 0.05$ &        $8.67 \pm 0.05$                  & $-0.92 \pm 0.13$ & \nodata & \nodata & \nodata & \nodata & \nodata & \nodata \\
NGC~185           &     10 /     440 &        $7.83 \pm 0.05$ &        $7.83 \pm 0.05$                  & $-1.12 \pm 0.36$ & \nodata & \nodata & \nodata & \nodata & \nodata & \nodata \\
NGC~147           & \phn 8 /     434 &        $7.79 \pm 0.05$ &        $8.00 \pm 0.05$                  & $-0.83 \pm 0.25$ & \nodata & \nodata & \nodata & \nodata & \nodata & \nodata \\
Andromeda~VII     & \phn 7 /     137 &        $7.22 \pm 0.13$ &        $7.17 \pm 0.13$                  & $-1.62 \pm 0.21$ & \nodata & \nodata & \nodata & \nodata & \nodata & \nodata \\
Andromeda~II      & \phn 9 / \phn 71 &        $6.96 \pm 0.08$ &        $6.96 \pm 0.08$                  & $-1.47 \pm 0.37$ & \nodata & \nodata & \nodata & \nodata & \nodata & \nodata \\
Andromeda~I       & \phn 2 / \phn 52 &        $6.68 \pm 0.05$ &        $6.88 \pm 0.05$                  & $-1.33 \pm 0.17$ & \nodata & \nodata & \nodata & \nodata & \nodata & \nodata \\
Andromeda~III     & \phn 3 / \phn 64 &        $6.00 \pm 0.12$ &        $6.26 \pm 0.12$                  & $-1.84 \pm 0.05$ & \nodata & \nodata & \nodata & \nodata & \nodata & \nodata \\
Andromeda~V       & \phn 4 / \phn 85 &        $5.75 \pm 0.09$ &        $5.79 \pm 0.09$                  & $-1.94 \pm 0.18$ & \nodata & \nodata & \nodata & \nodata & \nodata & \nodata \\
Andromeda~XVIII   & \phn 1 / \phn 18 &        $5.70 \pm 0.30$ &        $5.90 \pm 0.30$\tablenotemark{g} & $-1.35 \pm 0.20$ & \nodata & \nodata & \nodata & \nodata & \nodata & \nodata \\
Andromeda~XV      & \phn 1 / \phn 19 & $5.68^{+0.16}_{-0.13}$ & $5.89^{+0.16}_{-0.13}$\tablenotemark{g} & $-1.70 \pm 0.20$ & \nodata & \nodata & \nodata & \nodata & \nodata & \nodata \\
Andromeda~XIV     & \phn 2 / \phn 47 & $5.37^{+0.20}_{-0.30}$ & $5.58^{+0.20}_{-0.30}$\tablenotemark{g} & $-2.21 \pm 0.01$ & \nodata & \nodata & \nodata & \nodata & \nodata & \nodata \\
Andromeda~IX      & \phn 1 / \phn 32 &        $5.18 \pm 0.44$ &        $5.38 \pm 0.44$\tablenotemark{g} & $-1.93 \pm 0.20$ & \nodata & \nodata & \nodata & \nodata & \nodata & \nodata \\
Andromeda~X       & \phn 1 / \phn 27 &        $4.94 \pm 0.40$ &        $5.15 \pm 0.40$\tablenotemark{g} & $-2.46 \pm 0.20$ & \nodata & \nodata & \nodata & \nodata & \nodata & \nodata \\
\enddata
\tablerefs{The luminosities for most galaxies were taken from the June 2013 version of the compilation of \citet{mcc12} and references therein.  The luminosities and stellar masses for Canes Venatici~I through Coma Berenices were taken from \citet{mar08}.  Stellar masses for all other galaxies not marked with ``g'' or ``h'' were calculated by multiplying the luminosity by the stellar mass-to-light ratios of \citet{woo08}.}
\tablenotetext{a}{Number of member stars, confirmed by radial velocity, with measured [Fe/H].  For the M31 satellites, the number before the slash is the number of bins of coadded spectra, and the second number is the total number of individual spectra.}
\tablenotetext{b}{Mean [Fe/H] weighted by the inverse square of estimated measurement uncertainties.  We assume a solar abundance of $12 + \log ({\rm Fe/H}) = 7.52$.}
\tablenotetext{c}{The number in parentheses is the standard deviation of [Fe/H] weighted by the inverse square of the measurement uncertainties.}
\tablenotetext{d}{Median absolute deviation.}
\tablenotetext{e}{Interquartile range.}
\tablenotetext{f}{Actually the excess kurtosis, or 3 less than the raw kurtosis.  This quantifies the degree to which the distribution is more sharply peaked than a Gaussian.}
\tablenotetext{g}{Stellar mass-to-light ratio assumed to be $M_*/L_V = 1.6~M_{\sun}/L_{\sun}$, which is average value for dSphs measured by \citet{woo08}.}
\tablenotetext{h}{Stellar mass-to-light ratio assumed to be $M_*/L_V = 1.0~M_{\sun}/L_{\sun}$, which is average value for dTs measured by \citet{woo08}.}
\end{deluxetable*}

\begin{figure}[t!]
\centering
\includegraphics[width=\columnwidth]{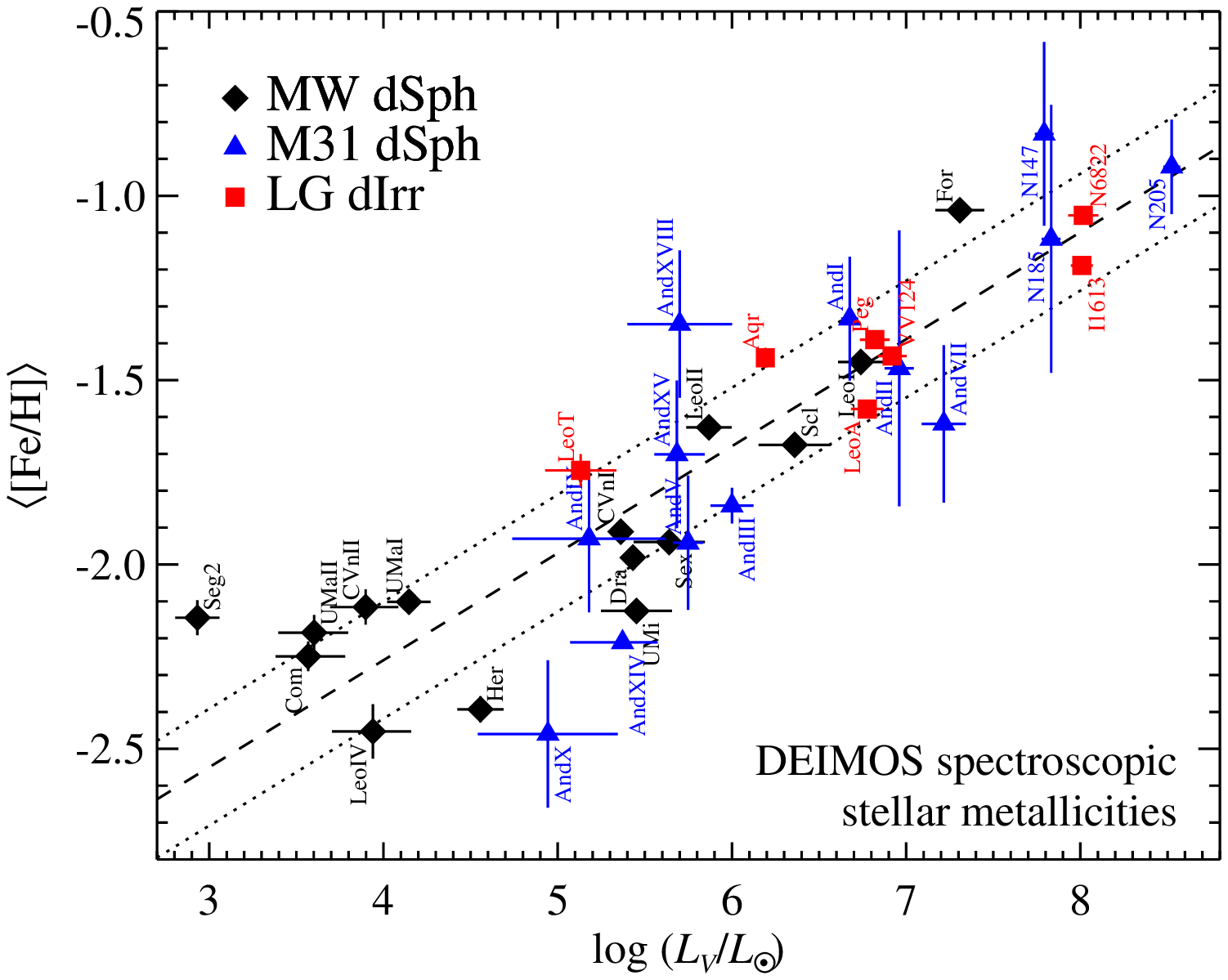}
\caption{The luminosity--stellar metallicity relation for Local Group
  dwarf galaxies.  The black diamonds (MW dSphs) and red squares
  (dIrrs) are the average stellar iron abundances from spectroscopy of
  individual stars.  The blue triangles (M31 dSphs) are the average
  stellar iron abundances from coadded spectroscopy of groups of
  similar stars within each dwarf galaxy.  The dashed line shows the
  least-squares line (Equation~\ref{eq:lzr}, where the intercept is
  calculated at $10^6~L_{\sun}$), excluding the M31 data points and
  Segue~2, which may be a heavily tidally stripped galaxy
  \citep{kir13}.  The dotted line shows the rms about the best fit.
  Unlike Figure~\ref{fig:gre03}, there are no photometric
  metallicities in this figure.  Hence, these data are not subject to
  the age--metallicity degeneracy.\label{fig:lzr}}
\end{figure}

\begin{figure*}[t!]
\centering
\includegraphics[width=\textwidth]{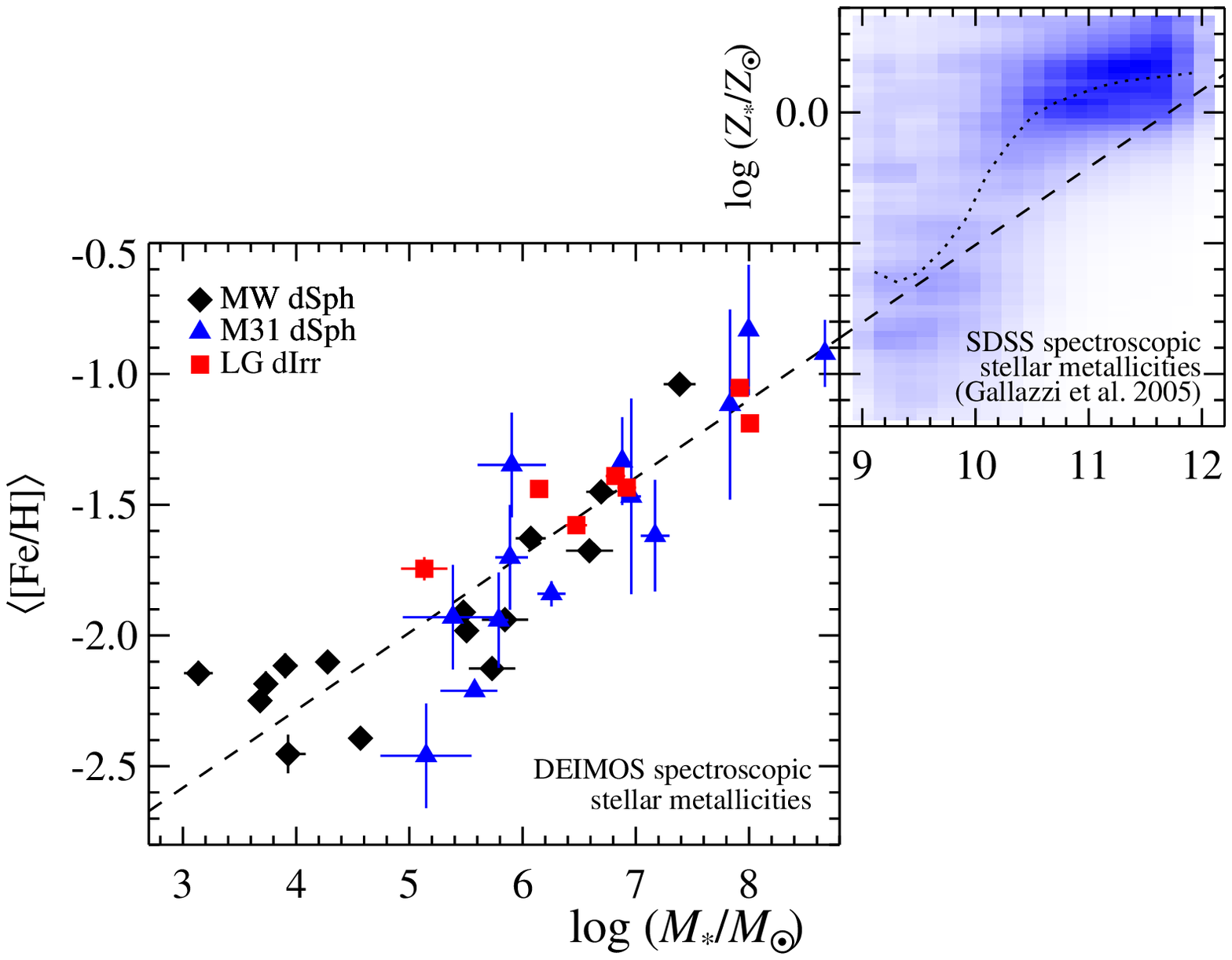}
\caption{The stellar mass--stellar metallicity relation for Local
  Group dwarf galaxies (left) and more massive SDSS galaxies galaxies
  \citep[right,][]{gal05}.  The Local Group metallicities ($\langle
  \mathfeh \rangle$) were measured from iron lines, and the SDSS
  metallicities ($\log Z_*$) were measured from a combination of
  absorption lines, mostly Mg and Fe.  The conversion between $\langle
  \mathfeh \rangle$ and $\log Z_*$ depends on [Mg/Fe].  The Local
  Group data is the same as in Figure~\ref{fig:lzr}, but it is plotted
  here versus stellar mass rather than luminosity.  The dashed line is
  the least-squares fit to the Local Group galaxies
  (Equation~\ref{eq:mzr}, where the intercept is calculated at
  $10^6~M_{\sun}$), and the dotted line in the right panel is the
  moving median for the SDSS galaxies.  Although the techniques at
  measuring both mass and metallicity differ between the two studies,
  the mass--metallicity relation is roughly continuous over nine
  orders of magnitude in stellar mass.\label{fig:mzr}}
\end{figure*}

The simplest diagnostic of differential chemical evolution among
galaxies is the LZR or MZR\@.  As discussed in
Section~\ref{sec:intro}, \citet{gre03} presented evidence that the LZR
for dwarf galaxies is dichotomous between dIrrs and dSphs (see
Figure~\ref{fig:gre03}).  However, all of their MW dSph metallicities
were based on spectroscopy whereas all of their dIrr metallicities
were based on broadband color.  Photometric metallicities are subject
to the age--metallicity degeneracy, which is difficult to resolve
without photometry reaching the main sequence turn-off.  Only recently
has such photometry become available for the dIrrs.  As \citet{lee08}
pointed out, the dichotomy in the LZR may be a result of the inability
to resolve the photometric age--metallicity degeneracy for dIrrs.
\citet{lia11} estimated that the effect of even a small (15\%)
intermediate-age population is a depression of $\mathfeh_{\rm phot}$
by a few tenths of a dex.

In order to address the dichotomy, we calculated average metallicities
from spectroscopy, which is not subject to the age--metallicity
degeneracy.  For each galaxy in Table~\ref{tab:sample}, we computed a
weighted mean of [Fe/H]\@.  The metallicity of each star in the
average was weighted by the inverse square of the error in [Fe/H]\@.
For the coadded spectra of M31 satellites, the average was computed
from bins of stars rather than individual stars.  For those galaxies
with only one bin (And IX, X, XV, and XVIII), the number presented as
$\langle \mathfeh \rangle$ is simply the metallicity of the bin.
Table~\ref{tab:lzr} and Figure~\ref{fig:lzr} show the resulting LZR,
separated by galaxy type (dSph or dIrr).

The LZR for dSphs and dIrrs is nearly identical.  The least-squares
fit for the MW dSphs, accounting for measurement uncertainty in both
$L_V$ and $\langle \mathfeh \rangle$ \citep{akr96}, is

\begin{equation}
\langle \mathfeh \rangle_{\rm dSph} = \left( \lfoffsetdsph \pm \lfoffseterrdsph \right) + \left( \lfslopedsph \pm \lfslopeerrdsph \right) \log \left(\frac{L_V}{10^6~L_{\sun}}\right) \: .  \label{eq:dsphlzr}
\end{equation}

\noindent
We excluded Segue~2 from this fit because it may be heavily tidally
stripped \citep{kir13}, and its present luminosity may not reflect its
luminosity when it finished forming stars.  We also excluded the M31
satellites because the technique used to measure their metallicities
is different and because their error bars are larger.  Including them
changes the slope and intercept by less than the uncertainties quoted
in Equation~\ref{eq:dsphlzr}.  The rms of the MW dSphs about
Equation~\ref{eq:dsphlzr} is $\lfrmsdsph$.

The LZR for the dIrrs is

\begin{equation}
\langle \mathfeh \rangle_{\rm dIrr} = \left( \lfoffsetdirr \pm \lfoffseterrdirr \right) + \left( \lfslopedirr \pm \lfslopeerrdirr \right) \log \left(\frac{L_V}{10^6~L_{\sun}}\right) \: . \label{eq:dirrlzr}
\end{equation}

\noindent
The rms of the dIrrs about Equation~\ref{eq:dirrlzr} is $\lfrmsdirr$.
The rms of the dIrrs about Equation~\ref{eq:dsphlzr} is
$\lfrmsdirrdsph$.  The dIrrs have a smaller scatter than dSphs about
the best-fit line for dSphs.

We conclude that dIrrs are not deviant from the LZR defined by MW
dSphs.  Both types of galaxies obey the same relation.  The
least-squares fit for the dIrrs and MW dSphs, again excluding Segue~2,
is

\begin{equation}
\langle \mathfeh \rangle = \left( \lfoffsetboth \pm \lfoffseterrboth \right) + \left( \lfslopeboth \pm \lfslopeerrboth \right) \log \left(\frac{L_V}{10^6~L_{\sun}}\right) \: . \label{eq:lzr}
\end{equation}

\noindent
The rms about the best-fit line is $\lfrms$.  Equation~\ref{eq:lzr} is
the dashed line in Figure~\ref{fig:lzr}.

Luminosity is a direct observable, but stellar mass is more closely
related to chemical evolution.  The mass-to-light ratio depends on the
SFH\@.  \citet{woo08} calculated $M_*/L_V$ for the brighter MW dSphs
and the LG dIrrs in two ways.  They used modeled SFHs \citep{mat98},
or they converted integrated galaxy colors into mass-to-light ratios
based on stellar population models \citep{bel01,bel03}.  Generally,
they preferred the SFH-based masses, but sometimes only integrated
colors were available.  For the fainter MW dSphs, we adopted
\citeauthor{mar08}'s (\citeyear{mar08}) stellar masses, which were
based on modeling the distribution of stars in the CMD for each
galaxy.  Table~\ref{tab:lzr} includes the stellar masses we adopted.
For those galaxies where stellar mass measurements were not available
from the aforementioned references, we assumed \citeauthor{woo08}'s
median SFH-based mass-to-light ratio for the appropriate galaxy type.
The footnotes in Table~\ref{tab:lzr} identify these galaxies.

In analogy to Figure~\ref{fig:lzr} for the LZR, Figure~\ref{fig:mzr}
shows the MZR for the same dwarf galaxies.  The least-squares fit
excluding Segue~2 and the M31 satellites is

\begin{equation}
\langle \mathfeh \rangle = \left( \mfoffsetboth \pm \mfoffseterrboth \right) + \left( \mfslopeboth \pm \mfslopeerrboth \right) \log \left(\frac{M_*}{10^6~M_{\sun}}\right) \: . \label{eq:mzr}
\end{equation}

\noindent
The rms about the best-fit line is $\mfrms$.  The scatter about the
MZR is about as small as the scatter about the LZR\@.  The similarity
is expected because the variance in $M_*/L_V$ is small---especially in
logarithmic space---for the predominantly old dwarf galaxies in the
LG\@.

Ignoring possibly tidally stripped galaxies like Segue~2, the MZR is
unbroken and of constant slope from the galaxies with the lowest known
stellar masses up to $M_* = 5 \times 10^8~M_{\sun}$ (NGC~205), which
is the upper stellar mass limit of our sample.  The continuity of the
relation begs the question, to what mass does the MZR persist?

The gas-phase MZR has been analyzed for many different galaxy masses,
SFRs, and redshifts.  Section~\ref{sec:intro} provides some background
on some of those studies.  However, the gas-phase metallicity depends
on the instantaneous SFR or gas fraction of the galaxy
\citep{man10,bot13}.  A more stable metallicity indicator is the
average metallicity, $\langle \mathfeh \rangle$, of the stars.
Besides, our measurements are of stellar metallicities.  After all,
the dSphs have no gas for which we could measure metallicity.
Therefore, it makes sense to compare our dwarf galaxy MZR to a study
of more massive galaxies with measurements of stellar metallicity.

\citet{gal05,gal06} compiled the largest sample of galactic stellar
metallicities.  They measured ages, metallicities, and an empirical
proxy for [$\alpha$/Fe] ratios for tens of thousands of SDSS galaxies.
The right panel of Figure~\ref{fig:mzr} shows a Hess diagram of their
MZR\@.  Although both our measurements and those of
\citeauthor{gal05}\ are stellar metallicities, they were not measured
in the same manner.  We measured [Fe/H] from iron absorption lines in
individual stars.  \citeauthor{gal05}\ measured metallicities using
broad spectral features, dominated by Mg and Fe, in the integrated
light of galaxies.  The $y$-axis labels in Figure~\ref{fig:mzr}
discriminate between the pure iron abundances and SDSS
``metallicities'' by calling them $\langle \mathfeh \rangle$ and $\log
(Z_*/Z_{\sun})$, respectively.  The conversion from $\langle \mathfeh
\rangle$ to $\log Z_*$ depends on [Mg/Fe].  For solar abundance
ratios, $\langle \mathfeh \rangle$ and $\log Z_*$ are directly
comparable.  The measurements also differ in many other ways, such as
the method of taking the average (averaging individual stars versus
light-weighted mean), the stellar population probed (red giants versus
the entire stellar population), and the models used
(\citeauthor{kir10}\ \citeyear{kir10}\ versus
\citeauthor{bru03}\ \citeyear{bru03}).

Despite the different techniques in measuring average metallicity and
stellar mass, these two samples are the best available to merge
together to form one MZR from the lowest to highest galaxy masses.
The dashed line in Figure~\ref{fig:mzr} is the best fit to the LG
dwarf galaxies (Equation~\ref{eq:mzr}).  The dotted line in the right
panel shows the moving median of the SDSS galaxies.  The shape of the
MZR for dwarf galaxies is a straight line in log--log space, but the
shape for the more massive galaxies flattens at higher metallicity.
The flattening also happens in the gas-phase MZR
\citep{tre04,and13,zah13}.  At least some of this flattening is due to
aperture bias of the SDSS fibers \citep[also see][]{kau03}.  The more
massive galaxies are rare and preferentially farther away.  The fixed
angular size of the fiber covers a larger fraction of these more
massive, more distant galaxies.  Radial metallicity gradients cause a
larger fraction of the outermost, metal-poor stars to be included in
the more massive galaxies.  In any case, the MZRs for the dwarf
galaxies and the SDSS galaxies are roughly continuous across the
boundary between the two samples of galaxies at $M_* = 10^9~M_{\sun}$.
The MZR slope is not quite the same at the $M_*$ boundary, but again,
the techniques at measuring stellar metallicities in the two samples
are not homogeneous.

A possible origin of the MZR is metal loss.  Galaxies with deeper
gravitational potential wells are able to retain supernova ejecta more
readily than less massive galaxies \citep[e.g.,][]{dek86}.  The less
massive galaxies lose gas and metals to supernova winds, and their
stars end up more metal-poor.  %The most massive galaxies attain a
%gravitational potential well depth high enough that supernova winds
%cannot blow out metals.  In other words, these galaxies become closed
%boxes.  In this context, the supernova feedback theory explains the
%turnover in the MZR at $\sim 10^{10.5}~M_{\sun}$.  The metallicities
%of more massive galaxies becomes independent of stellar mass because
%supernova winds cannot achieve the escape velocity.  Models of
%momentum-driven winds \citep{mur05} reproduce the shape of the MZR,
%including the turnover at high mass \citep{dav11}.

The success of feedback in explaining the MZR implies that dwarf
galaxies are extremely susceptible to metal loss.  \citet{kir11b}
showed that the MW satellite galaxies lost more than 96\% of the iron
that their stars produced.  That conclusion was based simply on the
stellar masses of the galaxies, theoretical nucleosynthetic yields,
and present stellar metallicities.  This metal loss could have been
caused by any gas loss mechanism, including supernova feedback
\citep[e.g.,][]{mur05}, radiation pressure \citep[e.g.,][]{mur11},
tidal stripping, or ram pressure stripping \citep[e.g.,][]{may01}.

The dIrr galaxies fall on the same MZR\@.  The same argument about
metal loss can be applied to them with one modification.  The MW dSphs
have no gas today, but dIrrs do have gas.  This difference alone
implicates gas stripping as a major cause for gas and metal removal
from the dSphs \citep[e.g.,][]{lin83}.  Any metals not present in
dIrrs' stars could be present in the gas.  \citet{gav13} proposed that
dIrrs could indeed evolve with no metal loss.  Instead, gas infall
combined with low star formation efficiencies (high gas mass
fractions) can be responsible for the low metallicities of dIrrs (also
see \citeauthor{mat94}\ \citeyear{mat94} and
\citeauthor{cal09}\ \citeyear{cal09}).  This scenario would lead to
high gas-phase metallicities.

However, the gas-phase metallicities of dIrrs are not especially high.
As an example, based on supernova yields \citep{iwa99,nom06} and a
Type~Ia supernova delay time distribution \citep{mao10}, the stellar
population in NGC~6822 should have produced $3 \times 10^5~M_{\sun}$
of iron.  Based on their average metallicity ($\langle \mathfeh
\rangle = \nsettfeh$), the stars in NGC~6822 harbor just 5\% of this
iron.  If the remaining iron were in the gas \citep[$M_{\rm HI} = 1.3
  \times 10^8~M_{\sun}$,][]{kor04}, then the metallicity of the gas
would be $\mathfeh = +0.1$.  The gas-phase oxygen
abundance\footnote{We assume a solar oxygen abundance of $12 + \log
  ({\rm O/H}) = 8.66$ \citep{asp04}.} is ${\rm [O/H]} = -0.55 \pm
0.10$ \citep{lee06b}.  Therefore, the gas would have ${\rm [O/Fe]} =
-0.7$ if the galaxy never lost its iron.  This value is at odds with
the stellar ratio \citep[${\rm [O/Fe]} = +0.1$,][]{ven01} measured
from young A supergiants.  Furthermore, the MZR shows no trend with
gas fraction.  DIrrs with gas-to-stellar mass ratios less than one,
like IC~1613 and VV~124, do not show any deviation from the MZR
compared to gas-rich dIrrs.  We conclude that the missing iron is no
longer part of stars or the star-forming gas.  Our present
measurements do not inform us whether the missing iron has left the
galaxy or has been incorporated into a hot gas halo
\citep{she12,she13}.

We have established that a single MZR applies to all LG galaxies with
$10^{3.5} < M_*/M_{\sun} < 10^9$, but we have not shown that all
galaxies in the universe obey such a tight relation.
\citet{gal05,gal06} separated more massive galaxies in the MZR into
high and low concentration groups.  The high-concentration galaxies
have lower SFRs than the low-concentration galaxies.  The two groups
also follow different MZRs.  The high-concentration galaxies have
higher metallicities on average, especially in the mass range $10^9 <
M_*/M_{\sun} < 10^{10.5}$.  The low-concentration galaxies have a
larger scatter in metallicity at fixed stellar mass.  The model of
\citet{mag12} explains these trends in the context of star formation
efficiency without invoking mass loss.  Denser galaxies (presumably
with denser gas) form stars more efficiently and end up with higher
metallicities.  Furthermore, satellite galaxies appear to be more
metal-rich at fixed halo or stellar mass than central galaxies
\citep{pas10}.  Therefore, there is a parameter other than $M_*$ that
controls the slope, offset, and scatter of the MZR for more massive
galaxies.

Our findings also cannot explain the offset in the MZR between dSphs
and dIrrs from abundances of PNe \citep{ric98,gon07}.  PNe in dSphs
are found to have very high oxygen abundances compared to their
stellar iron metallicities.  For example, the field population of
Fornax hosts one PN, which has an oxygen abundance between ${\rm
  [O/H]} = -0.7$ and $-0.3$ \citep{mar84,ric95,kni07}.  For
comparison, the stellar iron abundance is $\langle \mathfeh \rangle =
\forfeh$.  NGC~205 is another dSph/dE with very oxygen-rich PNe.
\citet{ric95} reported its mean PN abundance at ${\rm [O/H]} = -0.1$.
We measured its mean stellar iron abundance as $\langle \mathfeh
\rangle = \ntzffeh$.  Similarly, the PN abundances for NGC~147 and 185
are ${\rm [O/H]} = -0.6$ \citep{gon07} and $-0.5$ \citep{ric95},
respectively, whereas our stellar metallicities are $\langle \mathfeh
\rangle = \nofsfeh$ and $\noeffeh$.  It is possible that the PN
abundances are overestimated.  For example, \citeauthor{ric95}'s
average PN abundances were based on lower limits on the abundances for
several PNe in each galaxy.  As an example, the lower limits on
individual PNe in NGC~205 are up to 27 times smaller than the quoted
mean for the galaxy.  Perhaps the method of averaging lower limits
leads to a bias in the quoted abundance.  Alternatively, the dSph PNe
themselves might produce oxygen in the third dredge-up
\citep{mag05,kni07}.  The PNe might also be sampling a younger
population of stars that are preferentially more metal-rich than the
population average.  In the future, we will use measurements of
stellar magnesium abundances to compare to the oxygen abundances in
the PNe because oxygen is nucleosynthetically much more closely
related to magnesium than iron.

%%%%%%%%%%%%%%%%%%%%%%%%%%%%%%%%%
%%%%%%%%%   SECTION 5   %%%%%%%%%
%%%%%%%%%%%%%%%%%%%%%%%%%%%%%%%%%

\section{Galactic Chemical Evolution}
\label{sec:gce}

\tabletypesize{\scriptsize}

\begin{deluxetable*}{lcccccccccl}
\tablecolumns{11}
\tablewidth{0pt}
\tablecaption{Chemical Evolution Models\label{tab:gce}}
\tablehead{ & Leaky Box & & \multicolumn{3}{c}{Pre-Enriched} & & \multicolumn{3}{c}{Accretion} & \\ \cline{2-2} \cline{4-6} \cline{8-10}
\colhead{dSph} & \colhead{$p_{\rm eff}$\tablenotemark{a} ($Z_\sun$)} & & \colhead{$p_{\rm eff}$\tablenotemark{a} ($Z_\sun$)} & \colhead{${\rm [Fe/H]}_0$\tablenotemark{b}} & \colhead{$\Delta{\rm AICc}$\tablenotemark{c}}
& & \colhead{$p_{\rm eff}$\tablenotemark{a} ($Z_\sun$)} & \colhead{$M$\tablenotemark{d}} & \colhead{$\Delta{\rm AICc}$\tablenotemark{c}} & \colhead{Best Model}}
\startdata
\cutinhead{MW dSphs}
Fornax        & $0.106 \pm 0.005$         & & $0.082_{-0.004}^{+0.005}$ &             $-2.05 \pm 0.06$        &         \phs $124.03$ & & $0.111 \pm 0.003$         &    $ 9.3_{-1.3}^{+1.5}$ &          $306.90$ & Accretion    \\
Leo I         & $0.041 \pm 0.002$         & & $0.030 \pm 0.002$         &             $-2.33_{-0.06}^{+0.05}$ &         \phs $178.41$ & & $0.043 \pm 0.001$         &    $ 7.9_{-1.0}^{+1.2}$ &          $353.33$ & Accretion    \\
Sculptor      & $0.029 \pm 0.002$         & & $0.027 \pm 0.002$         &             $-3.39_{-0.26}^{+0.18}$ &     \phs\phn $ 10.72$ & & $0.029 \pm 0.002$         &    $ 1.4 \pm 0.2$       & \phn\phn $  5.32$ & Pre-Enriched \\
Leo II        & $0.028 \pm 0.002$         & & $0.024 \pm 0.002$         &             $-2.92_{-0.13}^{+0.11}$ &     \phs\phn $ 25.47$ & & $0.028 \pm 0.002$         &    $ 3.3_{-0.5}^{+0.7}$ &     \phn $ 45.22$ & Accretion    \\
Sextans       & $0.016 \pm 0.002$         & & $0.013 \pm 0.002$         &             $-3.17_{-0.23}^{+0.16}$ &     \phs\phn $ 12.01$ & & $0.014 \pm 0.001$         &    $ 3.3_{-1.0}^{+1.8}$ &     \phn $ 10.43$ & Pre-Enriched \\
Ursa Minor    & $0.011 \pm 0.001$         & & $0.007 \pm 0.001$         &             $-2.92_{-0.10}^{+0.09}$ &     \phs\phn $ 41.85$ & & $0.009 \pm 0.001$         &    $11.0_{-4.5}^{+5.6}$ &     \phn $ 44.30$ & Accretion    \\
Draco         & $0.014 \pm 0.001$         & & $0.011 \pm 0.001$         &             $-3.06_{-0.10}^{+0.09}$ &     \phs\phn $ 37.67$ & & $0.013 \pm 0.001$         &    $ 4.2_{-0.9}^{+1.3}$ &     \phn $ 44.70$ & Accretion    \\
Can.\ Ven.\ I & $0.019 \pm 0.002$         & & $0.016 \pm 0.002$         &             $-3.10_{-0.20}^{+0.15}$ &     \phs\phn $ 13.39$ & & $0.017_{-0.001}^{+0.002}$ &    $ 2.6_{-0.7}^{+1.0}$ & \phn\phn $  9.62$ & Pre-Enriched \\
\cutinhead{Local Group dIrrs}
NGC 6822      & $0.129 \pm 0.008$         & & $0.127_{-0.008}^{+0.009}$ &                          $<  -2.93$ &     \phn\phn $ -1.92$ & & $0.126_{-0.007}^{+0.008}$ &    $ 1.7_{-0.2}^{+0.3}$ & \phn\phn $  6.95$ & Accretion    \\
IC 1613       & $0.078_{-0.007}^{+0.008}$ & & $0.058_{-0.006}^{+0.007}$ &             $-2.08_{-0.11}^{+0.09}$ &     \phs\phn $ 29.82$ & & $0.075_{-0.005}^{+0.006}$ &    $ 4.3_{-1.1}^{+1.5}$ &     \phn $ 22.73$ & Pre-Enriched \\
VV 124        & $0.043_{-0.006}^{+0.007}$ & & $0.042_{-0.006}^{+0.007}$ &                          $<  -3.37$ &     \phn\phn $ -2.12$ & & $0.041 \pm 0.006$         &    $ 1.5_{-0.3}^{+0.7}$ &     \phn $ -2.63$ & Leaky Box    \\
Peg.\ dIrr    & $0.058_{-0.006}^{+0.007}$ & & $0.058_{-0.006}^{+0.007}$ &                          $<  -3.87$ &     \phn\phn $ -2.09$ & & $0.058 \pm 0.006$         &    $ 1.4_{-0.2}^{+0.4}$ &     \phn $ -1.76$ & Leaky Box    \\
Leo A         & $0.033 \pm 0.006$         & & $0.030_{-0.006}^{+0.007}$ &             $-3.06_{-2.88}^{+0.44}$ & \phs\phn\phn $  2.17$ & & $0.030 \pm 0.004$         &    $ 6.2_{-3.0}^{+4.1}$ & \phn\phn $  2.13$ & Pre-Enriched \\
Aquarius      & $0.044_{-0.009}^{+0.012}$ & & $0.039_{-0.010}^{+0.012}$ &                          $<  -2.08$ &     \phn\phn $ -0.11$ & & $0.040_{-0.006}^{+0.007}$ &    $ 5.6_{-2.7}^{+3.7}$ & \phn\phn $  2.87$ & Accretion    \\
Leo T         & $0.021_{-0.006}^{+0.008}$ & & $0.021_{-0.006}^{+0.007}$ &                          $<  -4.93$ &     \phn\phn $ -2.49$ & & $0.020 \pm 0.005$         &    $ 4.1_{-2.1}^{+6.3}$ &     \phn $ -2.12$ & Leaky Box    \\
\enddata
\tablecomments{Error bars represent 68\% confidence intervals.  Upper limits are at 95\% (2$\sigma$) confidence.}
\tablenotetext{a}{Effective yield.}
\tablenotetext{b}{Initial metallicity.}
\tablenotetext{c}{Difference in the corrected Akaike information criterion (Equation~\ref{eq:aicc}) between the specified model and the Leaky Box model.  Positive numbers indicate that the specified model is preferred over the Leaky Box model.}
\tablenotetext{d}{Accretion parameter, which is the ratio of final mass to initial gas mass.}
\end{deluxetable*}

\begin{figure*}[t!]
\centering
\includegraphics[width=\textwidth]{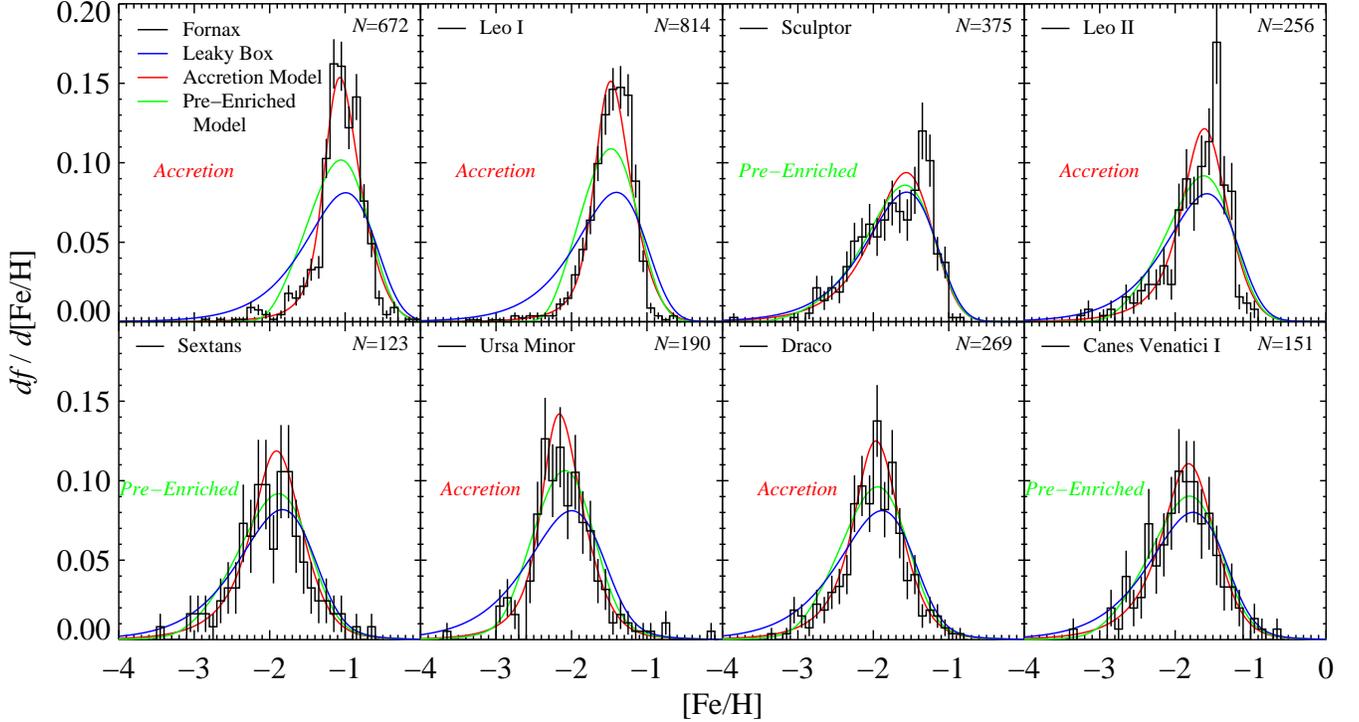}
\caption{The metallicity distributions for the eight most luminous MW
  dSphs in our sample.  The galaxies are arranged from most luminous
  (Fornax) to least luminous (Canes Venatici~I)\@.  The colored lines
  show the maximum likelihood fits of three different chemical
  evolution models convolved with the measurement uncertainties.  The
  colored, italicized text indicates the preferred
  model.\label{fig:dsph_feh_hist}}
\end{figure*}

\begin{figure*}[t!]
\centering
\includegraphics[width=\textwidth]{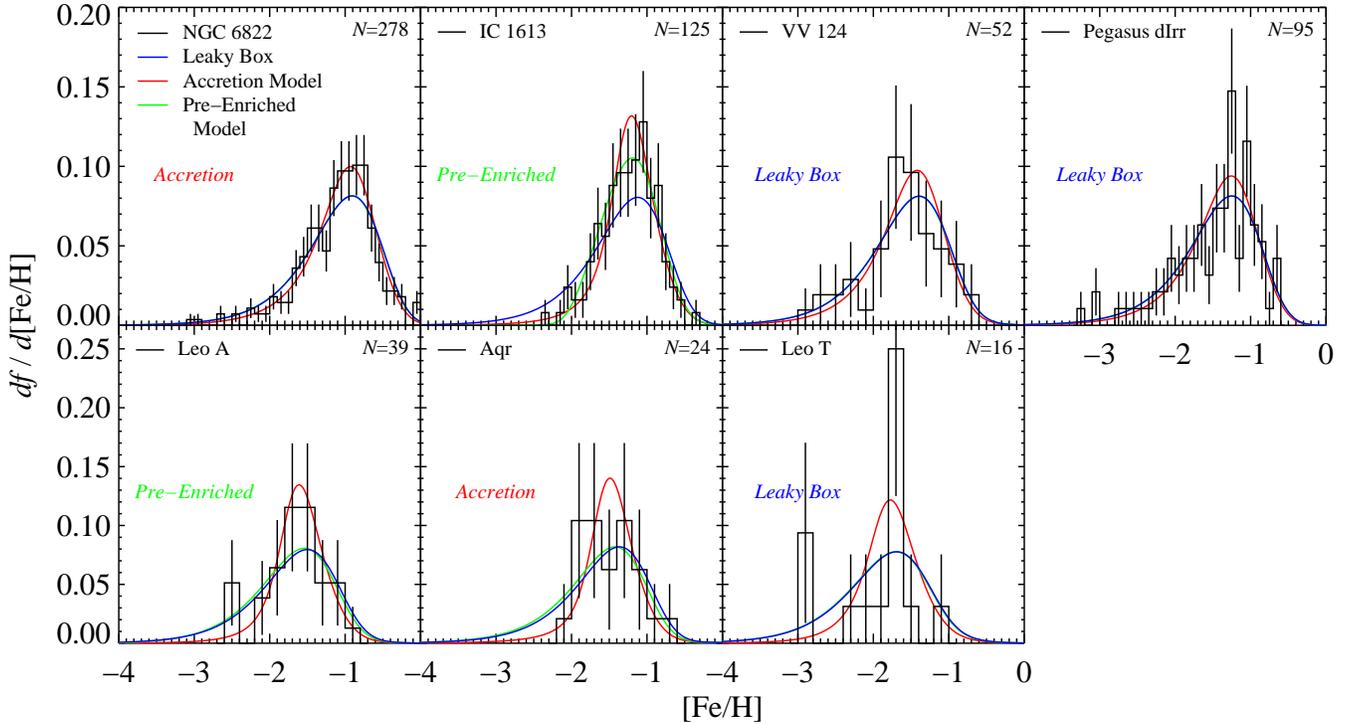}
\caption{The metallicity distributions for the dIrrs in our sample.
  The galaxies are arranged from most luminous (NGC~6822) to least
  luminous (Leo~T)\@.  Whereas the Accretion Model is generally a
  better description of dSphs (Figure~\ref{fig:dsph_feh_hist}), the
  Leaky Box or Pre-Enriched Models are fair descriptions of the
  dIrrs.\label{fig:dirr_feh_hist}}
\end{figure*}

\begin{figure}[t!]
\centering
\includegraphics[width=\columnwidth]{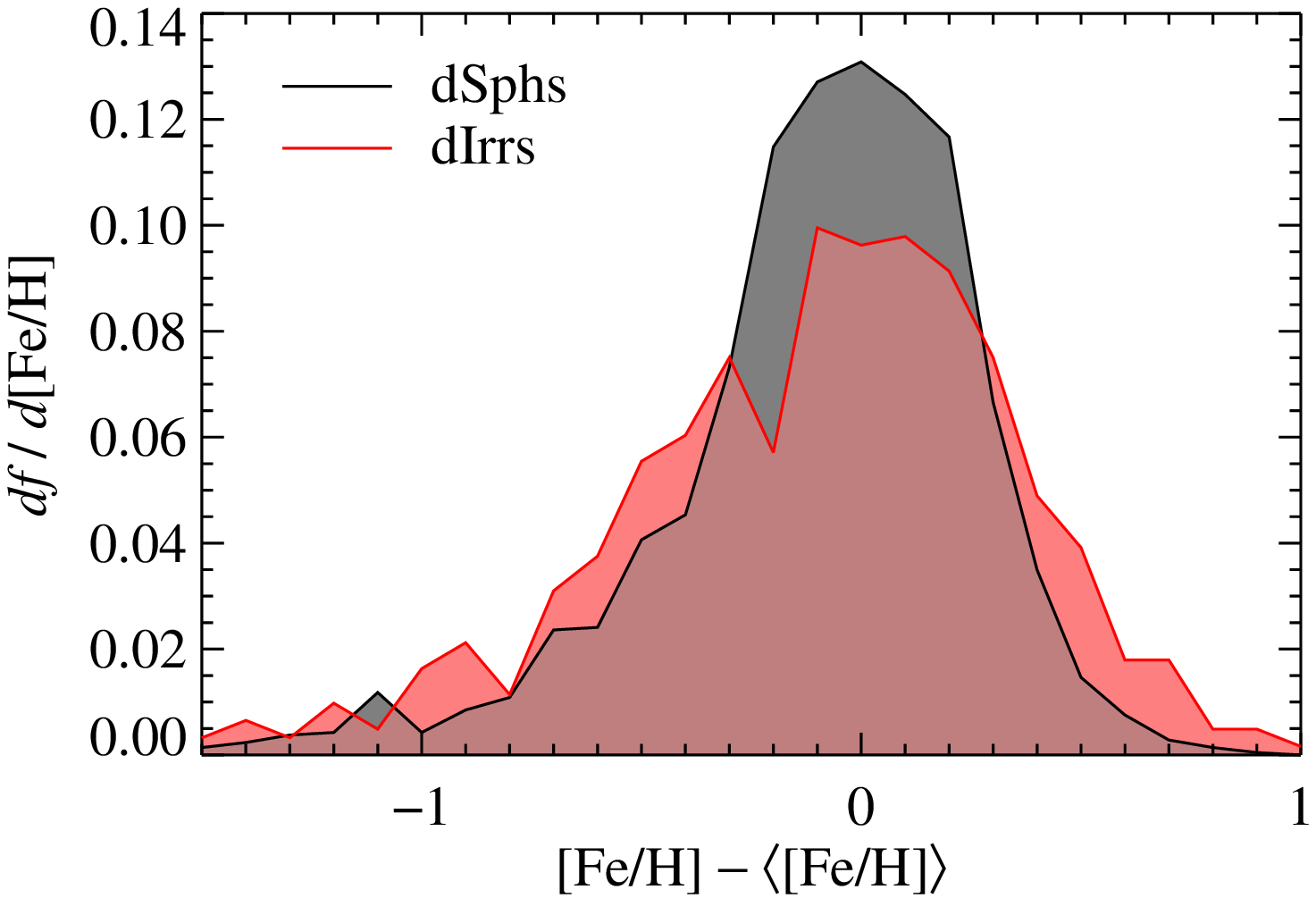}
\caption{The combined MDFs for the luminous dSphs in
  Figure~\ref{fig:dsph_feh_hist} (black) and the luminous dIrrs in
  Figure~\ref{fig:dirr_feh_hist} (red).  Only galaxies with $M_* >
  10^6~M_{\sun}$ were included in the stacked MDFs.  Each individual
  galaxy's MDF was centered at its mean [Fe/H] before the MDFs were
  stacked together.  The average MDF for dIrrs is broader and less
  peaked than the average MDF for dSphs.\label{fig:mdfshape}}
\end{figure}

Because we measured metallicities of individual stars in the dIrrs and
MW dSphs, we can analyze the metallicity distribution function (MDF)
rather than just the mean metallicity.  The MDF encodes the star
formation and gas flow history of the galaxy.  The shape of the MDF
indicates whether the galaxy conforms to a closed box or whether it
accreted gas during its star formation lifetime.
Figure~\ref{fig:dsph_feh_hist} shows the MDFs for the MW dSphs.  This
figure is nearly the same as Figure~1 of \citet{kir11a}.  We show it
here to contrast the dSphs with the dIrrs, whose MDFs are shown in
Figure~\ref{fig:dirr_feh_hist}.  Only stars with measurement
uncertainties $\delta\mathfeh < 0.5$ are included in those figures and
in the following discussion.

The MDFs of the dIrrs are shaped differently from the dSphs, even at
the same luminosity or stellar mass.  Three of the most luminous MW
dSphs---Fornax, Leo~I, and Leo~II---have narrowly peaked distributions
with a metal-poor tail.  Sculptor and the four least luminous MW dSphs
in Figure~\ref{fig:dsph_feh_hist} have broader MDFs.  The dIrr MDFs
are also broader even though six of the seven dIrrs have luminosities
similar to Leo~I and Fornax.  Figure~\ref{fig:mdfshape} illustrates
the different shapes.  The average MDF for the dIrrs with $M_* >
10^6~M_{\sun}$ is broader and less peaked than the average MDF for
dSphs in the same stellar mass range.  The two-sided
Kolmogorov--Smirnov test gives a probability of $\ksdsphdirr\%$ that
the distributions in Figure~\ref{fig:mdfshape} are drawn from the same
parent distribution.

Some of the difference in MDF shape between dSphs and dIrrs may
reflect the different SFHs between the two types of galaxies.  Because
dIrrs generally have more extended SFHs than dSphs
\citep{mat98,orb08}, they could have different [$\alpha$/Fe] ratios.
The dSph and dIrr MDFs of an $\alpha$ element, like oxygen or
magnesium, might be less diverse than the MDFs of iron.  Our spectral
synthesis technique of measuring abundances is sensitive to some
$\alpha$ elements.  In a future article, we will compare MDF shapes of
elements other than iron.

The uniformity of the MZR is even more remarkable in light of the
differently shaped MDFs.  Somehow, the mean metallicity of a dwarf
galaxy depends only on its stellar mass, regardless of how the
metallicities of individual stars are distributed about the mean.  We
return to this discussion in Section~\ref{sec:models}.

The shape of a galaxy's MDF can be understood in the context of its
history of star formation and gas flow.  For example, the accretion of
external, metal-poor gas can lower the metallicity of the galaxy's
star-forming interstellar medium (ISM)\@.  However, the presence of
new gas also triggers star formation, which raises the ISM
metallicity.  These effects can counteract each other to keep the ISM
metallicity roughly constant while stars are forming.  Therefore,
accretion of external gas can cause a peak in the MDF around a single
metallicity.

\subsection{Chemical Evolution Models}
\label{sec:models}

Quantitative models of galactic chemical evolution can be used to
interpret the shape of the MDF\@.  \citet{kir11a} fit three different
models of chemical evolution to the eight dSphs in
Figure~\ref{fig:dsph_feh_hist}.  We re-fit the same models to the dSph
MDFs, updated as described in
Section~\ref{sec:individualmetallicities}.  We also fit the same
models to the dIrrs.

All of the following analytic models assume the instantaneous mixing
and instantaneous recycling approximations.  The latter approximation
is not particularly appropriate for elements that have production
timescales delayed with respect to star formation.  For example, iron
is produced mostly in Type~Ia SNe, which are delayed with respect to
star formation.  It would be more appropriate to compare our models to
oxygen abundance distributions because oxygen production closely
tracks star formation.  However, iron is the best measured stellar
metallicity indicator available to us.  It must be kept in mind that
the instantaneous recycling approximation is a weakness in the
following models.

The simplest model is the Leaky Box \citep[called the Pristine Model
  by][]{kir11a}.  In this model, the galaxy begins its life with all
the gas it will ever have.  The gas is initially metal-free.  It may
turn into stars or be expelled from the galaxy.  The galaxy is not
allowed to accrete new gas.  The functional form of the Leaky Box is
the same as the Closed Box \citep{sch63,tal71,sea72}.  The only
difference is that the stellar nucleosynthetic yield ($p$) in the
Closed Box becomes the effective yield ($p_{\rm eff}$) in the Leaky
Box.  The definition of effective yield subsumes metal loss from the
galaxy.  The effective yield is the yield of metals that participate
in forming the next generation of stars.  The MDF of the Leaky Box is

\begin{equation}
\frac{dN}{d\mathfeh} \propto \left(\frac{10^{\mathfeh}}{p_{\rm eff}}\right) \exp \left(-\frac{10^{\mathfeh}}{p_{\rm eff}}\right) \: . \label{eq:leaky}
\end{equation}

\noindent
The only free parameter is $p_{\rm eff}$.

The Pre-Enriched Model \citep{pag97} is a generalization of the Leaky
Box, but the initial gas has a metallicity $\mathfeh_0$.  The MDF
of the Pre-Enriched Model is

\begin{equation}
\frac{dN}{d\mathfeh} \propto \left(\frac{10^{\mathfeh} - 10^{\mathfeh_0}}{p_{\rm eff}}\right) \exp \left(-\frac{10^{\mathfeh}}{p_{\rm eff}}\right) \: . \label{eq:preenriched}
\end{equation}

\noindent
The two free parameters are $p_{\rm eff}$ and $\mathfeh_0$.

The Accretion model \citep[called the Extra Gas Model by][]{kir11a} is
also a generalization of the Leaky Box.  \citet{lyn75} invented this
model and called it the Best Accretion Model.  The initial metallicity
is zero, but gas is allowed to flow into the galaxy according to a
specific functional form.  The MDF is described by two transcendental
equations that must be solved for the stellar mass fraction, $s$.

\begin{eqnarray}
\nonumber \mathfeh(s) &=& \log \bigg\{p_{\rm eff} \left(\frac{M}{1 + s - \frac{s}{M}}\right)^2 \times \\
            & & \left[\ln \frac{1}{1 - \frac{s}{M}} - \frac{s}{M} \left(1 - \frac{1}{M}\right)\right]\bigg\}\label{eq:s} \\
\nonumber \frac{dN}{d\mathfeh} &\propto&  \frac{10^{\mathfeh}}{p_{\rm eff}} \times \\
            & & \frac{1 + s\left(1 - \frac{1}{M}\right)}{\left(1 - \frac{s}{M}\right)^{-1} - 2 \left(1 - \frac{1}{M}\right) \times 10^{\mathfeh}/p_{\rm eff}}\label{eq:infall}
\end{eqnarray}

\noindent
The two free parameters are $p_{\rm eff}$ and the accretion parameter,
$M$, which is the ratio of the final mass to the initial gas mass.

We found the most likely parameters for all three models for all of
the dSphs in Figure~\ref{fig:dsph_feh_hist} and dIrrs in
Figure~\ref{fig:dirr_feh_hist}.  Following the same procedure as
\citet{kir11a}, we maximized the likelihood that the model described
the observed MDF with a Monte Carlo Markov chain.  The length of the
chain was $10^3$ trials for the Leaky Box and $10^5$ trials for the
Pre-Enriched and Accretion Models.  Table~\ref{tab:gce} gives those
parameters along with the 68\% likelihood intervals.
Figures~\ref{fig:dsph_feh_hist} and \ref{fig:dirr_feh_hist} show the
best-fitting model MDFs convolved with functions that approximate the
observational uncertainties for each galaxy.  Thus, the model curves
in Figure~\ref{fig:dirr_feh_hist} already reflect that the measurement
uncertainties are larger on average for the dIrrs compared to the
dSphs.

The Pre-Enriched and Accretion Models are generalizations of the Leaky
Box.  They always fit the MDF better than the Leaky Box because they
have two free parameters rather than one.  However, introducing a free
parameter into a model risks over-fitting the data.  The Bayesian
information criterion is a statistic that estimates whether the extra
free parameter is necessary.  A revision to this statistic is the
Akaike information criterion \citep[AIC,][]{aka74}.  \citet{sug78}
revised this statistic and called it the corrected AIC (AICc):

\begin{equation}
{\rm AICc} = -2 \ln L + 2r + \frac{2r(r+1)}{N-r-1}\label{eq:aicc}
\end{equation}

\noindent
where $L$ is the maximum likelihood of the model, $r$ is the number of
free parameters, and $N$ is the number of stars.  Table~\ref{tab:gce}
includes the value $\Delta {\rm AICc}$, which is the difference
between the AICc of the Pre-Enriched or Accretion Model and the Leaky
Box.  Positive values of $\Delta {\rm AICc}$ indicate that the
introduction of the extra free parameter is justified and that the
more complicated model fits better.  Negative values of $\Delta {\rm
  AICc}$ indicate that the extra free parameter is an unnecessary
complication.  The best model---the one with the largest AICc---is
indicated in Table~\ref{tab:gce} and in colored text in
Figures~\ref{fig:dsph_feh_hist} and \ref{fig:dirr_feh_hist}.

The Leaky Box is not the best model to describe any of the dSphs.
Five of the dSphs require the accretion of pristine gas to explain the
shapes of their MDFs.  Three of the dSphs are better described by the
Pre-Enriched Model.

On the other hand, the Leaky Box is the best model to describe the
MDFs of three of seven dIrrs.  The Accretion Model is the best model
only for NGC~6822 and Aquarius.  Even in those two cases, $\Delta {\rm
  AICc}$ is over six times smaller than $\Delta {\rm AICc}$ for any of
the dSph MDFs that prefer the Accretion Model.  In other words, those
two dIrrs prefer the Accretion Model, but not nearly as much as the
dSphs.

Again, the uniformity of the MZR is remarkable in light of the
different gas flow histories implied by the MDF shapes.  Despite the
varying importance of gas accretion and pre-enrichment, the mean
metallicity of dwarf galaxies is strictly a function of stellar mass.
The metallicity of any Closed Box galaxy approaches the
nucleosynthetic yield, regardless of stellar mass.  The average
metallicity of a Leaky Box is lower than the true yield only by virtue
of the expulsion of metals from the galaxy.  Therefore, the MZR can be
interpreted as a relation between stellar mass and metal loss.

However, the variable that controls metal loss is more likely to be
the depth of the gravitational potential well than stellar mass.
Therefore, the MZR may indicate that the stellar mass is an excellent
tracer of potential well depth.  Massive galaxies with deeper wells
retain more gas and hence form more stars.  Retention of gas goes hand
in hand with retention of metals produced by the stellar population.
The details of how the galaxy acquired the gas are not important.
%In essence, the gravitational potential alone is the origin of the
%MZR\@.

In support of this interpretation of the MZR, \citet{gal06} showed
that the average stellar metallicity of SDSS galaxies is a tighter
function of dynamical mass (essentially velocity dispersion) than
stellar mass (although we note that \citeauthor{tol11}
\citeyear{tol11} found that finding the total, virial masses is not
straightforward even for massive elliptical galaxies).  Unfortunately,
the velocity dispersion of dwarf galaxies traces only the innermost
mass.  The half-light radius is much smaller than the half-mass radius
for galaxies with $\sigma_v \la 10$~km~s$^{-1}$ \citep[see,
  e.g.,][]{wol10}.  Consequently, the dynamical masses of the dwarf
galaxies in our sample are virtually unconstrained compared to the
SDSS galaxies.  We cannot directly test the hypothesis that the
fundamental independent variable of the MZR is potential well depth
(virial mass) rather than stellar mass without assuming a strong
theoretical relation between inner mass and virial mass.

\subsection{Ram Pressure Stripping}

\begin{figure*}[t!]
\centering
\includegraphics[width=\textwidth]{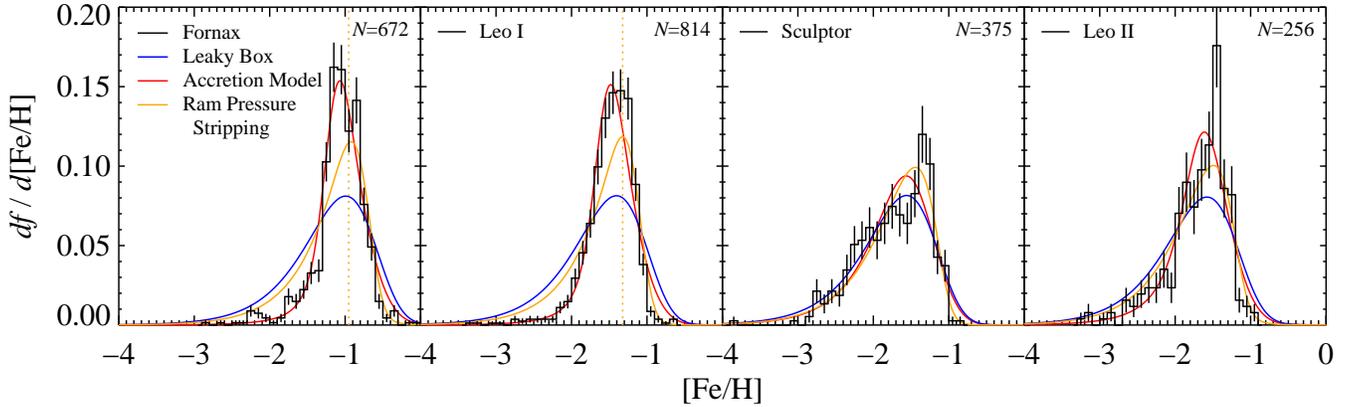}
\caption{The metallicity distributions for the four most luminous
  dSphs in our sample.  This figure is the same as the top row of
  Figure~\ref{fig:dsph_feh_hist} except that the Ram Pressure
  Stripping Model replaces Pre-Enriched Model.  The dotted orange line
  indicates $\mathfeh_s$, the metallicity at which ram pressure
  stripping turns on, in the two cases where it was able to be
  constrained.\label{fig:stripping}}
\end{figure*}

\begin{deluxetable}{lcccc}
\tablecolumns{5}
\tablewidth{0pt}
\tablecaption{Ram Pressure Stripping Model of Chemical Evolution\label{tab:stripping}}
\tablehead{ & \multicolumn{4}{c}{Ram Pressure Stripping} \\ \cline{2-5}
\colhead{dSph} & \colhead{$p_{\rm eff}$\tablenotemark{a} ($Z_\sun$)} & \colhead{${\rm [Fe/H]}_s$\tablenotemark{b}} & \colhead{$\zeta$\tablenotemark{c}} & \colhead{$\Delta{\rm AICc}$}}
\startdata
Fornax        & $0.465_{-0.101}^{+0.082}$ &             $-0.95 \pm 0.05$        & $2.01_{-0.70}^{+0.61}$ &         \phs $177.74$ \\
Leo I         & $0.178_{-0.033}^{+0.022}$ &             $-1.32_{-0.05}^{+0.04}$ & $2.30_{-0.65}^{+0.23}$ &         \phs $242.12$ \\
Sculptor      & $0.079_{-0.019}^{+0.048}$ &                          $<  -1.41$ & $0.43_{-0.17}^{+0.53}$ &     \phs\phn $ 21.88$ \\
Leo II        & $0.165_{-0.061}^{+0.027}$ &                          $<  -3.74$ & $1.68_{-0.82}^{+0.31}$ &     \phs\phn $ 43.26$ \\
\tablenotetext{a}{Effective yield.}
\tablenotetext{b}{Metallicity at which gas stripping commences.}
\tablenotetext{c}{Stripping parameter, which quantifies the rate of gas stripping.}
\enddata
\end{deluxetable}

None of the models is a great fit to the MDFs of the four most
luminous MW dSphs in our sample (Fornax, Leo~I, Sculptor, and
Leo~II)\@.  The observed MDFs approach a wall in [Fe/H] at the
metal-rich end.  None of the three models we presented so far can
explain the sharpness of the wall.  All three models overpredict the
number of the most metal-rich stars.  None of the dIrrs or the four
least luminous dSphs show this feature.

As the MW's satellite galaxies orbit around it, they pass through the
hot gas corona.  This corona can exert ram pressure on the galaxy's
gas.  Hydrodynamical models \citep[e.g.,][]{may06} show that ram
pressure stripping is effective at removing all of the gas from a
galaxy after just a couple pericentric passages.  In a new model
\citep{gat13} that incorporates supernova feedback, the MW removes all
of the gas from an infalling satellite galaxy in just one pericentric
passage.  The timescale for gas removal can be as short as 0.5~Gyr.

The LG galaxies show evidence for ram pressure stripping.
\citet{grc09} showed that nearly all galaxies within 270~kpc of the MW
or M31 have no gas.  Nearly all galaxies outside that boundary do have
gas.  Proximity to a large host galaxy is very effective at removing
gas.  \citeauthor{grc09}\ argued that the most likely culprit is ram
pressure stripping.

Rapid, efficient removal of gas can explain the sharp, metal-rich
cut-offs we observed for Fornax, Leo~I, Sculptor, and Leo~II\@.  A
simple modification to the Leaky Box model can predict the shape of
the metallicity distribution in the presence of ram pressure
stripping.\footnote{Tidal stripping or ram pressure stripping in
  conjunction with tidal stripping \citep{may01} can also remove gas.
  We call our model the Ram Pressure Stripping Model because it
  involves the rapid and terminal removal of gas.  Ram pressure
  stripping is more effective at that process than tidal stripping.}
This model is similar to the ``constant velocity flow'' model of
\citet{edm95}.  We assume that gas is removed at a constant rate
starting at time $t_s$.  Because the model has just one zone, time
$t_s$ corresponds to a metallicity $Z_s$.

At time $t=0$, the galaxy consists only of gas, and the gas mass
fraction is $g=1$.  The stellar mass fraction is $s$.  In the absence
of inflowing gas, the gas fraction is depleted by the outflow rate
($E$) and the SFR.

\begin{eqnarray}
\frac{dg}{dt} &=& -E - \frac{ds}{dt} \\
\frac{dg}{ds} &=& -\frac{E+ds/dt}{ds/dt} \label{eq:dgds}
\end{eqnarray}

\citet{pag97} derived the following relation between $g$, $s$,
metallicity ($Z$), and the nucleosynthetic yield ($p$) in the absence
of accretion.

\begin{equation}
g \frac{dZ}{ds} = p \label{eq:gdZds}
\end{equation}

\noindent
For convenience, we define $z \equiv Z / p$ such that $g \; dz/ds =
1$.  Combining Equations~\ref{eq:dgds} and \ref{eq:gdZds},

\begin{equation}
g \frac{dz}{dg} = -\frac{ds/dt}{E + ds/dt} \: . \label{eq:gdzds2}
\end{equation}

\noindent
For simplicity, we assume that the SFR is proportional to the gas
mass.

\begin{equation}
ds/dt = \beta g \label{eq:kslaw}
\end{equation}

\noindent
This is a simplified version of the Kennicutt--Schmidt law.

Now we assume that the gas outflow rate has a term proportional to the
SFR, such as would be the case with supernova feedback, and a constant
term that turns on after a time $t_s$, which mimics the commencement
of ram pressure stripping.

\begin{eqnarray}
E &=& \eta \frac{ds}{dt} + E_s' \label{eq:e} \\
E_s' &=& \left\{ \begin{array}{ll} \label{eq:es}
   0 & \mbox{ if $t < t_s$ or $z < z_s$} \\
   E_s & \mbox{ if $t \ge t_s$ or $z \ge z_s$}
\end{array} \right.
\end{eqnarray}

\noindent
Equation~\ref{eq:gdzds2} becomes

\begin{eqnarray}
g \frac{dz}{dg} &=& -\frac{\beta g}{(1+\eta)\beta g + E_s'} \: . \\
dz &=& -\frac{\beta \; dg}{(1+\eta)\beta g + E_s'} \\
e^{-(1+\eta)z} &\propto& (1+\eta)\beta g + E_s'
\end{eqnarray}

\noindent
We require that $g=1$ at $z=0$, and we require continuity in the
function $g$ at $t=t_s$.  These conditions combined with
Equation~\ref{eq:e} and \ref{eq:es} yield

\begin{equation}
g = \left\{ \begin{array}{ll}
  e^{-(1+\eta)z} & \mbox{ if $z < z_s$} \\
  \left[1 + \frac{E_s e^{(1+\eta)z_s}}{(1+\eta)\beta}\right]e^{-(1+\eta)z} - \frac{E_s}{(1+\eta)\beta} & \mbox{ if $z \ge z_s$}
\end{array} \right. \: .
\end{equation}

\noindent
For simplicity, we define a stripping parameter $\zeta \equiv
E_s/((1+\eta)\beta)$.  As for the Leaky Box model, an effective yield
can also be defined: $p_{\rm eff} \equiv p/(1+\eta)$ and $z \equiv
Z/p_{\rm eff}$.

The metallicity distribution in this model can be represented as
follows.

\begin{eqnarray}
\frac{ds}{d \log z} &=& (\ln 10) z \frac{ds}{dz} \\
 &=& (\ln 10)zg \\
 &=& \left\{ \begin{array}{ll}
  (\ln 10)ze^{-z} & \mbox{ if $z < z_s$} \\
  (\ln 10)z\left[e^{-z} + \zeta\left(e^{z_s-z} - 1\right)\right] & \mbox{ if $z \ge z_s$}
\end{array} \right.
\end{eqnarray}

\noindent
The first case ($z < z_s$) is identical to Equation~\ref{eq:leaky}.
The second case can be rewritten as

\begin{eqnarray}
\nonumber \frac{dN}{d\mathfeh} &\propto& \left(\frac{10^{\mathfeh}}{p_{\rm eff}}\right) \Bigg[ \exp \left(-\frac{10^{\mathfeh}}{p_{\rm eff}}\right) + \\
 & & \zeta \left( \exp \left(\frac{10^{\mathfeh_s}-10^{\mathfeh}}{p_{\rm eff}}\right) - 1\right) \Bigg] \: . \label{eq:stripping}
\end{eqnarray}

\noindent
The three free parameters are $p_{\rm eff}$, $\mathfeh_s$, and
$\zeta$\@.

We fit Equation~\ref{eq:stripping} to the MDFs of Fornax, Leo~I,
Sculptor, and Leo~II in the same manner that we fit the Pre-Enriched
and Accretion Models.  Figure~\ref{fig:stripping} shows the Ram
Pressure Stripping Model compared to the Leaky Box and Accretion
Models.  The dotted lines in the panels for Fornax and Leo~I show
$\mathfeh_s$.  In the cases of Sculptor and Leo~II, $\mathfeh_s$ is
very small.  In other words, ram pressure stripping turned on at the
same time that the galaxy started to form its first, most metal-poor
stars.  Table~\ref{tab:stripping} gives the best-fitting model
parameters as well as $\Delta {\rm AICc}$ compared to the Leaky Box.

The Ram Pressure Stripping Model fits all four dSphs better than the
Leaky Box, even after accounting for the addition of an extra two free
parameters.  In particular, the new model reproduces the sharpness of
the metal-rich cut-off.  In fact, the model was designed to do so.

However, the Ram Pressure Stripping Model is the best fit of all four
models only for Sculptor.  Fornax, Leo~I, and Leo~II prefer the
Accretion Model, although the difference between the two models for
Leo~II is tiny.  It is too simplistic to apply ram pressure stripping
to a Leaky Box to explain the MDF shapes of the dSphs.  It would be
better to apply ram pressure stripping to the Accretion Model to
explain the MDFs of Fornax and Leo~I\@.  We did not attempt to do so.
Rather our modification of the Leaky Box model already illustrates the
point that gas stripping can describe the metal-rich cut-offs.

The dIrrs and the four least luminous dSphs in our sample do not have
sharp metal-rich cut-offs.  The Ram Pressure Stripping Model is not
required to explain their MDFs.  It makes sense that dIrrs have not
encountered ram pressure stripping.  They are far from large galaxies
with gas halos that could strip them.  On the other hand, the
low-luminosity dSphs do orbit the MW at distances small enough to
encounter ram pressure stripping.  The fact that their MDFs show no
evidence for stripping may indicate that they finished their star
formation before they fell into the MW\@.  This interpretation is
consistent with the exclusively ancient populations in Sextans
\citep{orb08}, Ursa Minor \citep{mig99}, Draco \citep{gri98,apa01},
and Canes Venatici~I \citep{oka12}.  On the other hand, Fornax
\citep{col08} and Leo~I \citep{hel01} had SFHs easily extended enough
to be affected by ram pressure stripping.  While ram pressure
stripping likely ended star formation in Fornax and Leo~I, something
else ended star formation in the less luminous dSphs.  Suspects are
reionization---as long as it happened gradually enough to fail to
produce a metal-rich cut-off in the MDF---and supernova feedback.  We
note that \citet{mon10} and \citet{hid11} found no evidence for the
effects of reionization on the SFHs of isolated dwarf galaxies.

The differently shaped MDFs between dSphs and dIrrs pose a problem for
the theory that dIrrs transform into dSphs \citep{lin83,may01} unless
tidal stripping removed many metal-poor stars from dSphs as they fell
into the MW (see Section~\ref{sec:bias}).  Most of the dSphs have MDFs
that seem to require gas accretion.  Even incorporating an
environmental effect, like ram pressure stripping, cannot avoid the
fact that the MDF shapes of Fornax and Leo~I are not Leaky Boxes.
Sextans, Ursa Minor, Draco, and Canes Venatici~I are neither Leaky
Boxes nor ram pressure stripped.  Their MDF shapes are not that
simple.  On the other hand, the dIrr MDF shapes are fairly simple.
Even the dIrrs that prefer the Accretion Model have only a slight
preference (comparatively small values of $\Delta {\rm AICc}$).
Additionally, NGC~6822 has a low accretion parameter of $M = \nsettm$.
The accretion parameters for the dSphs that prefer the Accretion Model
range from $M = \leoiim$ (Leo~II) to $\umim$ (Ursa Minor).  In other
words, the dSph MDFs are not consistent with transforming the dIrr
MDFs via removal of gas associated with falling into the MW\@.

We have presented a limited set of chemical evolution models.  It is
possible that the Accretion Model fits the MDFs deceptively well.  The
physical interpretation of the model is not necessarily the truth of
the galaxy's history just because it fits the MDF\@.  More complex
models may better reflect the dSphs' SFH\@.  For example, Fornax
experienced multiple discrete episodes of star formation rather than
one smoothly varying SFR \citep{buo99,sav00,bat06,gul07,col08}.  A
proper chemical evolution model should incorporate multiple episodes
of star formation, the accompanying expulsion of gas, and the possible
subsequent re-accretion of gas (e.g., see the chemical evolution
models of \citeauthor{rom06}\ \citeyear{rom06} and
\citeauthor{yin10}\ \citeyear{yin10}).  It is also possible that
galaxies as large as Fornax were not completely ram pressure stripped
on their first pericentric approach to the MW\@.  After all,
simulations of tidal stirring \citep[e.g.,][]{may06,kaz11,lok11}
require multiple pericentric passages to complete the transformation
of a dIrr into a dSph.  It may be appropriate to add complexity to our
Ram Pressure Stripping Model to account for multiple pericentric
passages.  This is best achieved in hydrodynamical simulations
\citep[e.g.,][]{may06,gat13}.

%%%%%%%%%%%%%%%%%%%%%%%%%%%%%%%%%
%%%%%%%%%   SECTION 6   %%%%%%%%%
%%%%%%%%%%%%%%%%%%%%%%%%%%%%%%%%%

\section{Summary and Conclusions}
\label{sec:conclusions}

We measured metallicities from spectra of individual red giants in
15~MW dSphs and seven LG dIrrs as well as from coadded spectra of red
giants in 13~M31 dSphs.  In contrast to metallicities measured from
emission lines, our stellar metallicities were not affected by the
instantaneous gas fraction in the galaxies.  Instead, they are a
chronicle of the galaxies' past star formation.  Unlike integrated
light spectroscopy, resolving individual stars also allowed us to
explore the chemical evolution of an individual galaxy.  Our
measurements were based exclusively on \ion{Fe}{1} lines.  This
technique avoided some of the uncertainties associated with the
empirical calibration of CaT equivalent width and Lick indices.  While
these methods are sensitive to [Ca/Fe] and [Mg/Fe], respectively, our
method provides a direct measurement of iron abundance.

The stellar mass--stellar metallicity relation is roughly continuous
from the smallest galaxies ($M_* = 10^{3.5}~M_{\sun}$) to the largest
galaxies ($M_* = 10^{12}~M_{\sun}$).  The MZR measured from the
galaxies in our sample ($M_* < 10^9~M_{\sun}$) is $Z_* \propto
M_*^{\mfslopeboth \pm \mfslopeerrboth}$.  The slope of the MZR for
SDSS galaxies in the mass range $10^9 < M_*/M_{\sun} < 10^{10.5}$
\citep{gal05} is slightly steeper than for the dwarf galaxies, but the
comparison is approximate because the techniques used to measure
metallicities in the two stellar mass ranges were different.
%Above $M_* = 10^{10.5}~M_{\sun}$, the slope of the MZR decreases.

The MZR can be understood in the context of gas and metal flows.
Galaxies in less massive halos expel a larger fraction of their gas
because they lack the gravity to resist galactic winds.  The lost gas
carries away metals that the stellar population produced.  Without
those metals, the subsequent generations of stars are born more
metal-poor than they would have been in a more massive galaxy.  The
relationship between gravitational potential and metal retention is
especially apparent in the massive SDSS galaxies.  The stellar
metallicities in the SDSS galaxies follow a tighter relation with
potential well depth (dynamical mass) than stellar mass \citep{gal06}.
%Models of momentum-driven winds best reproduce the details of the
%shape and slope of the MZR \citep{dav11}.

An alternative interpretation of the MZR is variable star formation
efficiency.  Low mass galaxies have high gas mass fractions
\citep{beg08}.  Their gas dilutes the metals created by the stellar
population.  Consequently, the metallicity of the gas remains low, and
the stars that form from the gas are correspondingly metal-poor.
However, this explanation does not quite fit the gas fractions and
gas-phase metallicities of all galaxies.  In Section~\ref{sec:mzr}, we
showed that our measurement of NGC~6822's stellar metallicity implies
that 95\% of the iron created by the stellar population is missing
from the stars.  If the galaxy has not lost mass, then this iron must
be hiding in the gas.  Although gas-phase iron abundances are not
available, gas-phase oxygen abundance measurements combined with the
amount of missing iron imply an [O/Fe] ratio six times less than that
observed in stars.  We conclude that it is more likely that NGC~6822
is losing metals rather than harboring them in its gas.

DIrrs follow the same MZR as dSphs.  Photometric metallicities
previously indicated that the dIrrs are more metal-poor that dSphs at
fixed luminosity \citep{gre03}.  However, photometric metallicities
require a knowledge of the ages of stars.  The younger ages of dIrrs
compared to dSphs result in bluer red giants, which could be
interpreted as more metal-poor.  Our spectroscopic metallicities
circumvented the age--metallicity degeneracy, and we found no
significant difference in the average metallicities of dSphs and dIrrs
at fixed luminosity or stellar mass.

Despite the MZR's uniformity without regard to galaxy type, the
metallicity distributions of dSphs are different from dIrrs.  The
dSphs have narrow, peaked distributions compared to dIrrs.  All of the
dwarf galaxies have a tail of metal-poor stars, but this tail blends
more smoothly with the metal-rich stars for dIrrs than for dSphs.  The
MDFs of the four most luminous dSphs in our sample have sharp cut-offs
at high metallicity.

The shapes of the dIrrs' MDFs resemble a Leaky Box model of chemical
evolution.  Allowing for gas accretion improved the fit to the MDFs of
a couple dIrrs, like NGC~6822, but the amount of accretion required
was small.  On the other hand, most of the dSphs required a great deal
of gas accretion during their star formation lifetimes to explain the
shapes of their MDFs.  However, the chemical evolution models are
simplistic.  Most importantly, they assume the instantaneous recycling
approximation, which is not strictly applicable to our iron abundance
measurements.  The ideal model will both relax this approximation and
be set in a cosmological context.  In the meantime, our interpretation
of our model fits should be regarded as consistent with the data but
not a unique description.

The MDFs of the four most luminous dSphs in our sample show a sharp
cut-off at high metallicity.  The cut-off is nearly perfectly sharp
after accounting for observational uncertainty.  The less luminous
dSphs and the dIrrs do not show a cut-off.  The metallicity wall may
be environmental.  The luminous dSphs could still have been forming
stars when they fell into the hot gas corona of the MW\@.  The removal
of gas due to tidal or ram pressure stripping would have ended star
formation rapidly, leading to a sudden end to chemical evolution.
Conversely, the less luminous galaxies may have ended their star
formation due to reionization or internal mechanisms before they fell
into the MW\@.  As a result, their MDFs do not show a rapid cessation
in their chemical evolution.  The dIrrs have no metal-rich cut-off
because they are all too far from the MW or M31 to have experienced
ram pressure stripping.

Although tidal and ram pressure stripping are likely mechanisms for
metal loss, they do not shape the MZR\@.  Ram pressure stripping
freezes a galaxy's chemical evolution at the moment of infall, and
stripped galaxies still obey the MZR\@.  Therefore, galaxies obey the
MZR throughout their lives.

Despite widely varying SFHs, gas flow histories, and environments,
most galaxies in the universe adhere closely to the universal stellar
mass--stellar metallicity relation.  DSphs and dIrrs seem to be shaped
by environmental effects.  For example, they have very different gas
fractions \citep{grc09}, and even their metallicity distributions are
shaped differently.  Nonetheless, nearly all dwarf galaxies obey the
same MZR\@.  The relation indicates an inextricable connection between
the acquisition of stellar mass and the retention of metals.  The
processes that eject metals also expel gas that can no longer be used
to form stars.  These processes can be supernova feedback, stellar
winds, or ram pressure stripping.  Although the details of metal and
gas loss leave separate imprints on metallicity distributions, they
all preserve the universal stellar mass--stellar metallicity relation.

\acknowledgments We are grateful to the many people who have worked to
make the Keck Telescopes and their instruments a reality and to
operate and maintain the Keck Observatory.  The authors wish to extend
special thanks to those of Hawaiian ancestry on whose sacred mountain
we are privileged to be guests.  Without their generous hospitality,
none of the observations presented herein would have been possible.
We extend a special note of gratitude to Keck support astronomers Luca
Rizzi, Greg Wirth, and Marc Kassis.

We thank the anonymous referee for reviewing our manuscript.  We also
thank Yuichi Matsuda and Brenda Frye for obtaining the Subaru images
of Aquarius presented in Section~\ref{sec:aqr}, Andrew Cole and Mike
Irwin for sharing the INT Wide Field Survey photometric catalog of
Leo~A, Edouard Bernard for sharing his photometric catalog of IC~1613,
and Josh Simon and Marla Geha for sharing their DEIMOS spectroscopy of
Leo~T and other faint galaxies.  ENK thanks Evan Skillman, Leslie
Hunt, Laura Magrini, and Jose O{\~n}orbe for helpful conversations.
We thank Edouard Bernard for a careful reading of the manuscript and
helpful comments.  We also thank Namrata Anand for early contributions
to the coaddition of DEIMOS spectra.

ENK acknowledges support from the Southern California Center for
Galaxy Evolution, a multicampus research program funded by the
University of California Office of Research, and partial support from
NSF grant AST-1009973.  JGC thanks NSF grant AST-0908139 for partial
support.  PG acknowledges support from NSF grant AST-10-10039.  He
thanks the staff of the Aspen Center for Physics for their generous
hospitality during his visit.  LC was supported by UCSC's Science
Internship Program (SIP)\@.  AG acknowledges support from the EU
FP7/2007-2013 under grant agreement No. 267251 AstroFIt.  Funding for
the SDSS and SDSS-II has been provided by the Alfred P.\ Sloan
Foundation, the Participating Institutions, the National Science
Foundation, the U.S. Department of Energy, the National Aeronautics
and Space Administration, the Japanese Monbukagakusho, the Max Planck
Society, and the Higher Education Funding Council for England.  The
SDSS Web Site is \url{http://www.sdss.org/}.  IRAF is distributed by
the National Optical Astronomy Observatories, which are operated by
the Association of Universities for Research in Astronomy, Inc., under
cooperative agreement with the National Science Foundation.

%The SDSS is managed by the Astrophysical Research Consortium for the
%Participating Institutions. The Participating Institutions are the
%American Museum of Natural History, Astrophysical Institute Potsdam,
%University of Basel, University of Cambridge, Case Western Reserve
%University, University of Chicago, Drexel University, Fermilab, the
%Institute for Advanced Study, the Japan Participation Group, Johns
%Hopkins University, the Joint Institute for Nuclear Astrophysics, the
%Kavli Institute for Particle Astrophysics and Cosmology, the Korean
%Scientist Group, the Chinese Academy of Sciences (LAMOST), Los Alamos
%National Laboratory, the Max-Planck-Institute for Astronomy (MPIA),
%the Max-Planck-Institute for Astrophysics (MPA), New Mexico State
%University, Ohio State University, University of Pittsburgh,
%University of Portsmouth, Princeton University, the United States
%Naval Observatory, and the University of Washington.  {\it Facility:}

\facility{Keck:II (DEIMOS)}

\clearpage
\renewcommand{\thetable}{\arabic{table}}
\setcounter{table}{2}
\begin{turnpage}
\tabletypesize{\scriptsize}
\begin{deluxetable}{llcccccccccccc}
%\tabletypesize{\scriptsize}
\tablewidth{0pt}
\tablecolumns{14}
\tablecaption{Metallicity Catalog\label{tab:catalog}}
\tablehead{\colhead{Galaxy} & \colhead{Name} & \colhead{RA (J2000)} & \colhead{Dec (J2000)} & \colhead{Filter 1} & \colhead{Magnitude 1\tablenotemark{a}} & \colhead{Filter 2} & \colhead{Magnitude 2\tablenotemark{a}} & \colhead{Filter 3} & \colhead{Magnitude 3\tablenotemark{a}} & \colhead{$T_{\rm eff}$} & \colhead{$\log g$} & \colhead{$\xi$} & \colhead{[Fe/H]} \\
\colhead{ } & \colhead{ } & \colhead{ } & \colhead{ } & \colhead{ } & \colhead{(mag)} & \colhead{ } & \colhead{(mag)} & \colhead{ } & \colhead{(mag)} & \colhead{(K)} & \colhead{(cm~s$^{-2}$)} & \colhead{(km~s$^{-1}$)} & \colhead{(dex)}}
\startdata
NGC~6822 & M07--9512  & 19 44 25.92 & $-$14 50 30.2 & $B$ & $22.337 \pm 0.210$ & $V$ & $21.878 \pm 0.058$ & $I$ & $20.718 \pm 0.008$ & $4485 \pm  57$ & $1.19 \pm 0.05$ & $1.86$ & $-0.56 \pm 0.11$ \\
NGC~6822 & M07--9518  & 19 44 25.94 & $-$14 52 05.7 & $B$ & $22.627 \pm 0.101$ & $V$ & $21.640 \pm 0.032$ & $I$ & $20.366 \pm 0.001$ & $4305 \pm  32$ & $0.97 \pm 0.04$ & $1.91$ & $-2.23 \pm 0.17$ \\
NGC~6822 & M07--9630  & 19 44 26.35 & $-$14 50 49.8 & $B$ & $22.314 \pm 0.065$ & $V$ & $21.207 \pm 0.023$ & $I$ & $19.916 \pm 0.001$ & $4335 \pm  28$ & $0.80 \pm 0.09$ & $1.95$ & $-1.20 \pm 0.11$ \\
NGC~6822 & M07--9787  & 19 44 26.98 & $-$14 52 02.5 & $B$ & $23.010 \pm 0.116$ & $V$ & $21.397 \pm 0.027$ & $I$ & $19.808 \pm 0.001$ & $3957 \pm  15$ & $0.55 \pm 0.05$ & $2.01$ & $-2.46 \pm 0.12$ \\
NGC~6822 & M07--9903  & 19 44 27.44 & $-$14 49 42.4 & $B$ & $22.972 \pm 0.159$ & $V$ & $22.292 \pm 0.084$ & $I$ & $21.187 \pm 0.001$ & $4540 \pm  90$ & $1.41 \pm 0.06$ & $1.81$ & $-0.90 \pm 0.12$ \\
NGC~6822 & M07--10027 & 19 44 27.91 & $-$14 50 08.3 & $B$ & $21.756 \pm 0.045$ & $V$ & $21.172 \pm 0.020$ & $I$ & $20.020 \pm 0.001$ & $4478 \pm  34$ & $0.94 \pm 0.04$ & $1.92$ & $-0.39 \pm 0.11$ \\
NGC~6822 & M07--10044 & 19 44 27.95 & $-$14 51 21.5 & $B$ & $22.717 \pm 0.110$ & $V$ & $21.815 \pm 0.043$ & $I$ & $20.304 \pm 0.001$ & $4031 \pm  22$ & $0.81 \pm 0.13$ & $1.95$ & $-1.53 \pm 0.12$ \\
NGC~6822 & M07--10225 & 19 44 28.45 & $-$14 50 42.3 & $B$ & $22.854 \pm 0.194$ & $V$ & $21.778 \pm 0.063$ & $I$ & $20.329 \pm 0.008$ & $4084 \pm  34$ & $0.85 \pm 0.02$ & $1.94$ & $-2.00 \pm 0.12$ \\
NGC~6822 & M07--10380 & 19 44 28.94 & $-$14 52 39.5 & $B$ & $22.075 \pm 0.139$ & $V$ & $21.519 \pm 0.029$ & $I$ & $20.175 \pm 0.001$ & $4239 \pm  25$ & $0.85 \pm 0.04$ & $1.94$ & $-1.76 \pm 0.12$ \\
NGC~6822 & M07--10421 & 19 44 29.05 & $-$14 50 07.6 & $B$ & $21.985 \pm 0.068$ & $V$ & $21.731 \pm 0.051$ & $I$ & $21.074 \pm 0.001$ & $5688 \pm 117$ & $1.72 \pm 0.02$ & $1.73$ & $-0.44 \pm 0.16$ \\
\nodata & \nodata & \nodata & \nodata & \nodata & \nodata & \nodata & \nodata & \nodata & \nodata & \nodata & \nodata & \nodata & \nodata \\
\enddata
%\tablerefs{In the online version of this paper, photometry is given
%  from the following sources: Sculptor, \citet{wes06}; Fornax,
%  \citet{ste98}; Leo~I, \citet{soh07}; Sextans, \citet{lee03}; Leo~II,
%  unpublished, but obtained in the same fashion as for Leo~I; Canes
%  Venatici~I, SDSS DR5 \citet{ade07}; Ursa Minor, \citet{bel02}; and
%  Draco, \citet{seg07}.}
\tablenotetext{a}{Corrected for extinction.}
\tablecomments{Table~\ref{tab:catalog} is published in its entirety in
  the electronic edition of the Astrophysical Journal.}  %Some columns
%  (right ascension, declination, and $B$, $V$, $R$, and $I$ magnitudes)
%  are suppressed in the printed edition.}
\end{deluxetable}
\clearpage
\end{turnpage}

\end{document}